\documentclass[aps,prd,twocolumn,showpacs,10pt,superscriptaddress,preprintnumbers,nofootinbib]{revtex4-1}
\usepackage{epsfig,amssymb,amsmath,psfrag,epstopdf,color}
\pdfoutput=1
\usepackage{graphicx}
\usepackage{multirow} 
\usepackage{makecell}
\usepackage{subcaption}
\usepackage{ragged2e}
\DeclareCaptionJustification{justified}{\justifying}
 \usepackage[normalem]{ulem}
\usepackage{paralist}
\usepackage{hyperref}[colorlinks=true,linkcolor=blue,citecolor=blue,urlcolor=blue]
\usepackage[size=tiny]{todonotes}
\usepackage{xspace}
\usepackage{cleveref}

\def\beq{\begin{equation}}
\def\eeq{\end{equation}}
\def\bsp#1\esp{\begin{split}#1nd{split}}
\newcommand{\be}{\begin{equation}}
\newcommand{\ee}{\end{equation}}
\newcommand{\bea}{\begin{eqnarray}}
\newcommand{\eea}{\end{eqnarray}}

\def\to{\rightarrow}

\def\td{\mathrm{d}}

\begin{document}
%%%%%%%%%%%%%%%%%%%%%%%%%%%%%%%%%%%%%%%%%%
\preprint{MIT-CTP/5982}

%%%%%%%%%%%%%%%%%%%%%%%%%%%%%%%%%%%%%%%%%%
\title{Living on the edge: radius effects in the angular substructure of heavy-ion jets}
%%%%%%%%%%%%%%%%%%%%%%%%%%%%%%%%%%%%%%%%%%

%%%%%%%%%%%%%%%%%%%%%%%%%%%%%%%%%%%%%%%%%%
%%%%%%%%%%%%%%%%%%%%%%%%%%%%%%%%%%%%%%%%%%

\author{Carlota Andres}
\email{carlota.andres@polytechnique.edu}
\affiliation{CPHT, CNRS, Ecole Polytechnique, Institut Polytechnique de Paris, 91120 Palaiseau, France}
\affiliation{Center for Theoretical Physics - a Leinweber Institute, Massachusetts Institute of Technology, Cambridge, MA 02139, USA}

\author{Jack Holguin}
\email{jack.holguin@manchester.ac.uk}
\affiliation{Department of Physics \& Astronomy, University of Manchester, Manchester M13 9PL, United Kingdom}

\author{Benjamin Kimelman}
\email{benjamin.kimelman@vanderbilt.edu}
\affiliation{Department of Physics and Astronomy, Vanderbilt University, Nashville, TN}

\author{Raghav Kunnawalkam Elayavalli}
\email{rithya.ke@vanderbilt.edu}
\affiliation{Department of Physics and Astronomy, Vanderbilt University, Nashville, TN}

\author{Jussi Viinikainen}
\email{jussi.viinikainen@vanderbilt.edu}
\affiliation{Department of Physics and Astronomy, Vanderbilt University, Nashville, TN}

\author{Zhong Yang}
\email{zhong.yang@vanderbilt.edu}
\affiliation{Department of Physics and Astronomy, Vanderbilt University, Nashville, TN}

%%%%%%%%%%%%%%%%%%%%%%%%%%%%%%%%%%%%%%%%%%
%%%%%%%%%%%%%%%%%%%%%%%%%%%%%%%%%%%%%%%%%%

\begin{abstract}
Jet substructure observables serve as essential tools for probing the quark-gluon plasma  produced in relativistic heavy-ion collisions. Their  interpretation, however, is often complicated by \emph{edge effects}, which arise when correlated particles fall outside the reconstructed jet radius, introducing  distortions that obscure the underlying QCD dynamics. In this work, we present a comprehensive phenomenological study of edge effects in soft-insensitive angular observables, taking the two-point energy correlator (EEC) as a representative example. We argue that these distortions scale linearly with the average angular separation between the winner-take-all and $E$-scheme axes $\langle \phi \rangle$, and validate this behavior across proton-proton (p-p) simulations with \textsc{Pythia}8 and \textsc{Herwig}7, as well as  lead-lead (Pb-Pb) simulations using \textsc{JEWEL} and \textsc{CoLBT}. In p-p collisions, edge effects are strongly suppressed, scaling as $(R_L/R)^4$, whereas medium-modified jets can exhibit larger distortions, with contributions scaling as $(R_L/R)^2$ and $(R_L/R)^4$. Taking Pb-Pb/p-p ratios of the EEC  substantially reduces, but does not completely eliminate, these distortions, highlighting the need of accounting for edge effects in the interpretation of heavy-ion jet substructure measurements. Since edge effects are largely governed by the $\langle \phi \rangle$ distribution, studying this distribution provides a new handle for benchmarking and constraining the modeling of edge effects in heavy-ion  event generators.
\end{abstract}

%%%%%%%%%%%%%%%%%%%%%%%%%%%%%%%%%%%%%%%%%%
%%%%%%%%%%%%%%%%%%%%%%%%%%%%%%%%%%%%%%%%%%
\maketitle

\section{Introduction}
\label{sec:intro}

High-energy jets offer a powerful window into the droplets of quark-gluon plasma (QGP) produced in heavy-ion collisions at the Relativistic Heavy Ion Collider (RHIC) and at the Large Hadron Collider (LHC) \cite{Busza:2018rrf,Harris:2023tti,Cao:2020wlm,Cunqueiro:2021wls,Apolinario:2022vzg,ALICE:2022wpn,CMS:2024krd}. In recent years, growing attention has been devoted to the theoretical modeling and experimental exploration of the inner structure of jets, commonly referred to as jet substructure \cite{Dasgupta:2013ihk,Larkoski:2017jix,Marzani:2019hun}, which provides unique avenues to study the microscopic dynamics of the QGP \cite{Cao:2020wlm,Cunqueiro:2021wls,Apolinario:2022vzg,ALICE:2022wpn,CMS:2024krd}. Among various jet substructure observables, angular measurements in heavy-ion collisions, such as the differential jet shape \cite{CMS:2013lhm,CMS:2018zze,CMS:2018jco,CMS:2021nhn,CMS:2022btc}, the groomed jet radius \cite{ALargeIonColliderExperiment:2021mqf,ATLAS:2022vii,CMS:2024zjn} and energy correlators \cite{CMS:2025ydi,talk_ananya} show clear deviations from their proton-proton (p-p) baseline, particularly at large angles within the jet cone. Interpreting these angular observables, however, requires careful consideration of \emph{edge effects}.

Edge effects refer to any distortions in jet substructure observables that stem from the jet reconstruction procedure itself, such as the choice of jet algorithm or jet radius, and occur in any collision system. They appear when particles correlated with the jet fall outside the reconstructed jet cone and are thus excluded from the measurement. Edge effects influence all classes of jet substructure observables and, in the worst-case scenario, may cause them to reflect artifacts of the reconstruction procedure rather than the genuine QCD dynamics of the jet core. As a result, multiple independent substructure measurements---each designed to probe different aspects of jet core dynamics---may instead become a highly correlated measurement of the jet clustering, undermining their effectiveness as independent probes of underlying QCD processes. While edge effects affect both angular and non-angular substructure observables, they are expected to be more tractable in the former because their impact is typically localized near the boundary of the jet cone. For this reason, our focus will be on edge effects in angular observables.

\begin{figure}
\includegraphics[scale=0.47]{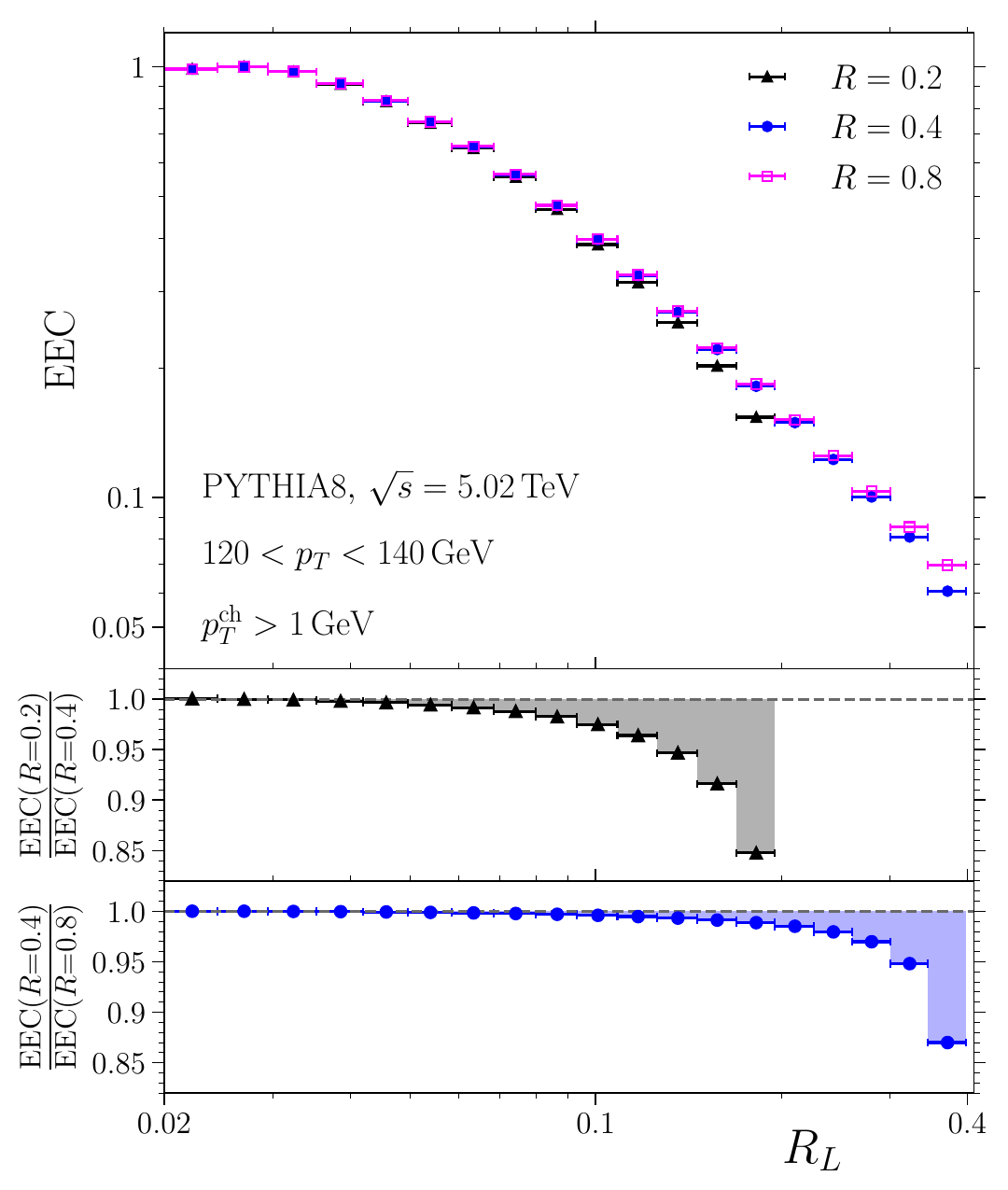}
\caption{Top: two-point energy correlator for anti-$k_T$ jets with $120 < p_T < 140$~GeV, computed using the same event sample in $\sqrt{s}=5.02$~TeV p-p collisions, for jet radii $R = 0.2$ (black triangles), $R = 0.4$ (blue circles), and $R = 0.8$ (pink squares). Middle: Ratio between the EEC for $R = 0.2$ and $R = 0.4$. Bottom: Ratio between the EEC for $R = 0.4$ and $R = 0.8$.} 
\label{fig1}
\end{figure}

To illustrate the concept of edge effects, let us consider an angular jet substructure measurement performed by Tahani on a sample of jets clustered with a jet radius $R=0.2$. Later, Chidi reconstructs the same observable using the same sample but with a larger radius  $R=0.4$. Central to the paradigm of jet substructure is the expectation that Chidi's and Tahani's measurements should reveal qualitatively and quantitatively similar features within the $R=0.2$ jet core, providing a consistent picture of  the underlying dynamics. Notable differences between their measurements  may indicate that excluding wider-angle hadrons in Tahani's jets introduces significant distortions. Similarly, Chidi’s measurement could be affected by the exclusion of hadrons at even larger angles, beyond $R=0.4$, which fall outside the reconstructed jets. These potential discrepancies are a manifestation of edge effects.  This is exemplified for a specific angular jet substructure observable, the two-point energy correlator (EEC) \cite{Basham:1978bw,Basham:1978zq,Basham:1979gh}  in Fig.~\ref{fig1},  where $R_L$ is the angle between the two detectors. Tahani’s $R=0.2$ analysis (black triangles) and Chidi’s $R=0.4$ result  (blue circles) are obtained from the same sample of events in $\sqrt{s}=5.02$ TeV p-p collisions generated with \textsc{Pythia} 8.230 \cite{Sjostrand:2014zea}. Differences between the two stem solely from edge effects, which become pronounced for $R_L \gtrsim 0.1$, as shown in the middle-panel ratio. The black-shaded area helps to visualize these missed correlations in Tahani's measurement relative to Chidi's. Naturally, when Eleanor sees these results they become concerned about possible edge effects in Chidi’s $R=0.4$ measurement and repeats the analysis using an $R=0.8$ jet radius, confirming that Chidi's result measurement is also affected by correlations missed out due to the imposition of the $R=0.4$  jet radius, as illustrated as the blue-shaded area in the bottom panel.

Our expectation that the measurements from Tahani, Chidi, and Eleanor should capture the same underlying physics in the jet core is grounded in the factorization of dynamics across different scales. Clustering effects become significant when hadrons have angular separations comparable to the jet radius, whereas the internal jet dynamics is typically dominated by its hard-collinear core. Because all three analyses use the same set of events, the physics governing the core should remain insensitive to the specific clustering algorithm or jet radius, provided that the chosen radius is sufficiently large. \footnote{We note that selection bias does not affect our heavy-ion results, because we only compare each heavy-ion sample to itself (not to a p–p sample).}

In heavy-ion collisions, edge effects pose a particular pressing challenge because QGP-induced modifications are most pronounced at large angles \cite{Connors:2017ptx,Cao:2020wlm,Cunqueiro:2021wls,Apolinario:2022vzg,ALICE:2022wpn,CMS:2024krd,CMS:2013lhm,CMS:2018zze,CMS:2018jco,CMS:2021nhn,CMS:2022btc,ALargeIonColliderExperiment:2021mqf,ATLAS:2022vii,CMS:2024zjn,CMS:2025ydi}, precisely the regime where angular observables are  most vulnerable  to such distortions. For instance,  recent measurements of the EEC within inclusive jets in lead-lead (Pb-Pb) collisions at the LHC  show clear deviations from their p-p baseline  for angular separations $R_L \gtrsim 0.1$ \cite{CMS:2025ydi,talk_ananya}. Notably, this angular region coincides with the domain where edge effects distort the observable in both p-p (as illustrated in Fig.~\ref{fig1}) and Pb-Pb (to be shown in section~\ref{sec:MC}). This overlap naturally raises a critical question: to what extent do the observed modifications in Pb-Pb reflect genuine QGP-induced phenomena, and how much might be attributed to reconstruction artifacts?

Although edge effects are a generic feature of jet substructure observables, our interest in them has been driven by the study of energy correlators (for a review on these observables, see \cite{Moult:2025nhu}). This focus does not reflect a particular vulnerability of energy correlators to such distortions; rather, it stems from the rare opportunity for analytical control they offer. Thanks to the light-ray operator product expansion (OPE), energy correlators are more amenable for an analytical treatment than traditional jet substructure tools \cite{Hofman:2008ar,Korchemsky:2019nzm,Dixon:2019uzg,Lee:2022uwt,Chen:2023zlx,Andres:2024xvk}. Yet, these calculations inherently neglect effects arising from practical aspects of jet clustering and background subtraction. To fully exploit the theoretical advantages of energy correlators, it is therefore essential to control or mitigate edge effects. Moreover, accurate modeling of these effects in Monte Carlo simulations is also critical for disentangling genuine QGP-induced jet modifications from algorithmic artifacts in comparisons to experimental data.

Motivated by these considerations, our study addresses two key questions:
\begin{enumerate}
    \item How significant are edge effects in angular jet substructure observables in heavy-ion collisions?
    \item Can these edge effects be mitigated to enable more robust theory-data comparisons?
\end{enumerate}
We use the two-point energy correlator, an increasingly popular observable \cite{Andres:2022ovj,Andres:2023xwr,Yang:2023dwc,Barata:2023bhh,Bossi:2024qho,Fu:2024pic,Andres:2024hdd,Xing:2024yrb,Singh:2024vwb,Apolinario:2025vtx,Barata:2025fzd,ALICE:2024dfl,CMS:2024mlf,CMS:2025ydi,talk_ananya,STAR:2025jut},  as a case study to address these questions. Our findings naturally extend to other soft-insensitive angular jet substructure observables that share a similar factorization structure, such as the differential jet shape $\rho(r)$ \cite{CDF:2005prv,CMS:2013lhm,CMS:2018zze,CMS:2018jco,CMS:2021nhn,CMS:2022btc} and the soft-drop groomed jet radius $\theta_g$ with $\beta = 0$ \cite{Larkoski:2014wba,ALargeIonColliderExperiment:2021mqf,CMS:2024zjn,ATLAS:2022vii}. For soft-sensitive observables, such as angularities, we expect qualitatively similar behavior; however, the factorization framework presented in this work does not apply directly and would require a dedicated generalization. 

We begin in  Section~\ref{sec:factorization} by reviewing the factorization structure of the EEC, which provides the foundation for understanding  edge effects in angular jet substructure.  In Section~\ref{sec:analytical}, we introduce a simplified analytical model, within which we isolate the leading contribution to these edge effects. We find that, for any collision system, the leading contribution is governed by the mean decorrelation between the jet axis defined in the $E$-scheme and the winner-takes-all (WTA) axis, denoted by $\langle \phi \rangle$.  
The main body of this paper, presented in Section~\ref{sec:MC}, systematically tests these analytical insights through event-generator simulations in p-p using \textsc{Pythia8} and \textsc{Herwig7}, and in Pb-Pb with \textsc{CoLBT} and \textsc{JEWEL}. Finally, in Section~\ref{sec:conclusions}, we discuss the broader implications of these findings, particularly for achieving more precise (analytical) theory-experiment comparisons in jet angular substructure studies and for improving how edge effects are modeled in heavy-ion event generators.

\section{Analytical framework}
\label{sec:factorization}

Measurements of energy correlations among the constituents of inclusive jet samples can be described analytically using a leading-power factorization theorem \cite{Dixon:2019uzg,Lee:2024icn,Lee:2024tzc}. This factorization has been rigorously established for vacuum jets, where a systematic framework for incorporating higher order corrections is also well understood \cite{Manohar:2002fd,Bauer:2003mga,Schindler:2023cww,Lee:2024esz,Chen:2024nyc}. More recently, it has been argued that the same factorization may be extended to describe energy correlator measurements within high-energy jets propagating through a colored medium \cite{Andres:2023xwr,Andres:2024ksi,Andres:2024xvk,Singh:2024vwb}. However, in this case, the systematic inclusion of subleading effects remains an open question.

Before presenting the full mathematical formulation of this factorization, let us  provide an intuitive overview of the key ingredients. The EEC factorization exploits the hierarchy among three physical scales: the initial hard partonic transverse momentum $p_{T}'$, the transverse momentum within the jet, $\sim R p_T'$, and the transverse momentum between hadrons involved in the EEC measurement,  approximately given by $R_L p_T'$ where $R_L$ is the correlator angle. This scale separation leads to a factorized expression consisting of three main components. First, the \emph{hard function} $H(p_{T}')$ encodes the production of the high transverse momentum parton that initiates the jet. Next, the \emph{jet function} $\mathcal{J}(R p_T')$ describes the evolution of this hard parton into a jet of hadrons, reconstructed using a sequential recombination algorithm with jet radius $R$. Finally, the \emph{EEC jet function} $j(R_L p_T')$ encodes the EEC measurement performed on the jet constituents. The $p_{T}$-spectrum of the jets on which the EEC measurement is performed is described by a DGLAP-style convolution of $H$ and $\mathcal{J}$, while the function $j$ captures the  angular energy correlation within jets of that given $p_{T}$. Returning to the example of Tahani, Chidi and Eleanor, within this factorized picture, their measurements differ only by $\mathcal{J}(R p_T')$, which does not depend directly on the measured EEC, and thus cannot describe an $R_L /R$-dependent suppression due to edge effects.

This factorization structure is universal for projected energy correlators within jets, including those with energy weights larger than one \cite{Dixon:2019uzg, Chen:2020vvp,Lee:2025okn}. The main difference between two-point and higher-point projected correlators lies in the specific calculation of the EEC jet function. For the two-point correlator, the jet and EEC jet functions are known analytically in vacuum at two loops \cite{Dixon:2019uzg,Chen:2023zlx,Lee:2024tzc}, with evolution kernels available up to three loops \cite{Mitov:2006ic,Moch:2007tx,Chen:2020uvt}. To reduce notational complexity, we focus here on the simplest case: the $n=1$ energy weighted two-point correlator measured on jets produced at central rapidities in hadron collisions. In this case, the transverse momentum of a jet hadron approximates its energy, and $\Delta R_{ij} = \sqrt{\Delta \eta_{ij}^2 + \Delta \phi_{ij}^2}$ approximates the angle $\theta_{ij}$ between the two jet hadrons three-momenta. Extension to  jets at moderately forward rapidities involves a straightforward modification to make the factorization differential in the jet pesudorapidity \cite{Kang:2016mcy,Singh:2024vwb}. 

Within this setup, the two-point energy correlator inside an inclusive sample of jets at central rapidities is given by
\begin{align}
    &\frac{\td \sigma_{\rm EEC}(R)}{\td p_T \td R_L } = \nonumber \\
    &\sum_{i,j} \int \frac{\td \sigma_{ij}(R)}{\td p_T^{i,j} \, \td \Delta R_{ij} \, \td p_T} \, \frac{p_T^i p_T^j}{ p_T^2 } \delta(R_L -\Delta R_{ij}) \, \td p_T^{i,j} \, \td \Delta R_{ij}\,,
\end{align}
where $\td \sigma_{ij}(R)$ is the inclusive differential cross section to produce hadrons $i,j$  with transverse momenta $p_T^i$ and $p_T^j$ within a jet of radius $R$ and transverse momentum $p_T$. The factorization theorem takes the form \cite{Kang:2016mcy,Dixon:2019uzg,Lee:2024icn} 
\begin{align}
     &\frac{\td \sigma_{\rm EEC}(R)}{\td p_T \td R_L } = \int \td p_{T}' \int \frac{\td x}{x} \,H_a \left( x , \frac{p_T'^2}{\mu^2} \right) \delta(p_T - x p_T')  \nonumber \\ & \times \int  \td y \, y^2   \mathcal{J}_{ab} \left(p_T', y, \ln\! \frac{x^2 R^2 p_T'^2}{\mu^2} \right) 
     \frac{j_b\left(\ln \! \frac{y^2 R_L^2 p_T'^2}{\mu^2} \right) }{R_L } \nonumber \\ \,& \times \left(  1 + \mathcal{O}\left(\frac{R_L^2 }{R^2}\right)\right)\,, \label{eq:refactorised}
\end{align}
where $p_{T}'$ is the transverse momentum of the initial hard parton, the indices $a$ and $b$ run over parton flavors, and $\mu$ denotes the renormalization scale. Although each individual function on the right-hand side of \eqref{eq:refactorised} depends explicitly on $\mu$, the full expression is independent of $\mu$ when computed at all orders, as required for a physical cross section. Logarithmic resummation is achieved by solving the renormalization group equations and choosing  $\mu$ appropriately  for each function, while ensuring that the overall dependence cancels.

As emphasized above, this factorization is valid in the collinear regime of the EEC, where the the transverse momentum between hadrons is much smaller than the jet transverse momentum, implying $R_L \ll R$. The $\mathcal{O}(R_L^2 /R^2)$ term in \eqref{eq:refactorised} explicitly indicates that the leading corrections to the collinear limit scale maximally with $R_L^2$. This relatively mild scaling ensures that this factorization holds well even at moderate values of $R_L$. For a perturbative evaluation of the EEC distribution, one would additionally require $R_L \gg \Lambda_{\rm QCD}/p_T$.  However, this condition is not necessary  for our present analysis.

At leading logarithmic (LL) accuracy, the EEC spectrum can be obtained by evaluating the hard function and jet functions at tree level.\footnote{We emphasize that LL accuracy for the EEC is single-log accuracy. When compared with typical Sudakov observables, this accuracy is equivalent to next-to-leading log.} This corresponds to the replacements
\begin{align}
   \mathcal{J}_{ab} \left(p_T', y, \ln\! \frac{x^2 R^2 Q^2}{\mu^2} \right) &\mapsto \mathcal{J}_{ab} \left(p_T', \ln\! \frac{x^2 R^2 Q^2}{\mu^2} \right) \delta(y-1), \nonumber \\
    H_a\left( x , \frac{p_T'^2}{\mu^2} \right) &\mapsto H_a\left( \frac{p_T'^2}{\mu^2} \right) \delta(x-1)\,,
\end{align}
which eliminate the convolution integrals in Eq.~\eqref{eq:refactorised}.  The resulting LL-accurate expression for the EEC spectrum becomes
\begin{align}
    \frac{\td \sigma_{\rm EEC}^{ \rm LL}(R)}{\td p_T \td R_L } = 
     H_a \left(\frac{p_T^2}{\mu^2} \right) \mathcal{J}_{ab} \left(p_T, \ln\! \frac{R^2 p_T^2}{\mu^2} \right) \frac{j_b\left(\ln \! \frac{ R_L^2 p_T^2}{\mu^2} \right)}{R_L }  \,. \label{eq:factorization_LL}
\end{align}
In this LL factorized picture, it becomes evident that the dependence on the jet algorithm enters exclusively through the  prefactor $\mathcal{J}$, which depends on  $p_T$ and $R$ but not on $R_L$. As a result, in the collinear regime $R_L \ll R$ where the factorization is valid, the angular dependence of the EEC is fully governed  by the function $j$, and thus remains unaffected by the choice of the jet algorithm encoded in $\mathcal{J}$. Furthermore, the prefactor $\mathcal{J}$ is constrained by the sum rule,
\begin{equation}
    \int \td R_L \, \frac{\td \sigma}{\td p_T \td R_L } = \frac{\td \sigma}{\td p_T} \,,
\end{equation}
removing the need of knowing $\mathcal{J}$ for computing the EEC at LL. In addition, the LL prefactor $\mathcal{J}$ cancels in ratios of measurements performed at the same jet $p_T$ and $R$, such as those commonly used when comparing results from heavy-ion and p-p collisions. Therefore, a detailed understanding of $\mathcal{J}$ is only necessary for high-precision studies, beyond the accuracy typically achieved in current heavy-ion jet analyses.

Such a factorized approach has played a central role in establishing jet substructure observables as precision tools in p-p collisions. Before incorporating edge effects into this picture, it is instructive to highlight why this framework can become particularly powerful for energy correlator measurements in heavy-ion environments. In the factorization of a general observable, the relevant matrix elements, encoded in the hard and jet functions, and, when needed, beam and soft functions, completely determine the observable. However, evaluating these matrix elements in the presence of a colored medium remains a significant challenge, despite recent progress \cite{Caron-Huot:2010qjx,Blaizot:2012fh,Apolinario:2014csa,Sievert:2018imd,Feal:2018sml,Blagojevic:2018nve,Mehtar-Tani:2019tvy,Sievert:2019cwq,Andres:2020vxs,Attems:2022ubu,Isaksen:2023nlr,Andres:2023jao,Soudi:2024yfy,Mehtar-Tani:2025rty}.  Energy correlators offer a distinct advantage over other jet substructure observables because they allow a higher degree of analytical control over both the EEC jet functions $j$ and the hard function $H$, even in the presence of a finite colored medium. This control arises from the light-ray OPE~\cite{Hofman:2008ar, Andres:2024xvk},  which systematically organizes contributions in the collinear limit and provides an alternative route to compute the jet and hard function matrix elements. Nevertheless, edge effects, arising at $R_L \lesssim R$, may compromise the predictive power of this factorized framework  and therefore must be carefully understood.

In any collision system, edge effects generally induce an $R_L$-dependent suppression of the EEC spectrum, becoming significant when $R_L \lesssim R$, as some particles correlated with the jet fall outside the jet radius---see, for instance, Fig.~\ref{fig1}. These losses can be captured by a more complex factorization theorem that involves only two ingredients. The first is the same hard function $H$ as in the original  factorization. The second is  a new \textit{combined jet function} $\mathcal{G}$, which describes both the jet clustering and the EEC measurement. This simultaneous description is significantly more challenging than the  separate treatment of the functions $\mathcal{J}$ and $j$. In particular, $\mathcal{G}$ is defined by multi-scale matrix elements whose structure is intricate and cannot be tackled using the light-ray OPE. 

The resulting factorization in terms of the combined jet function takes the form 
\begin{align} 
\frac{\td \sigma_{\rm EEC}(R)}{\td p_T \td R_L } = \int &\frac{\td x}{x}\, \delta(p_T - x p_T')\, H_a \left( x , \frac{p_T'^2}{\mu^2} \right)\nonumber \\ & \times \, \mathcal{G}_{a} \left(p'_T, R_L , R , \ln\! \frac{x^2 R^2 p_T'^2}{\mu^2} \right) \, ,
\label{eq:factorization2}
\end{align}
where, notably, $\mathcal{G}_a$ depends on both $R$ and $R_L$. Importantly, the breakdown of the simpler factorized picture, as in Eq.~\eqref{eq:refactorised}, is not unique to energy correlators; it applies to all jet substructure observables sensitive to emissions near the edge of the jet, although the detailed  behavior of these effects depends on the observable.

With this in mind, our goal is to answer the initial questions formulated in Section~\ref{sec:intro}. In particular, we aim to determine the size of the leading corrections to Eq.~\eqref{eq:refactorised} arising from edge effects, and to assess whether these leading contributions are similar in Pb-Pb and p-p collisions.  Once this is established, we seek to explore whether it is possible to mitigate or control the leading edge effects in heavy-ion jet substructure measurements, thereby enabling a proper quantification of genuine QGP-induced modifications in theory-data comparisons. 

Addressing these questions for heavy-ion jets using a fully analytical approach is not currently possible, even at leading-log accuracy. Doing so would require computing $\mathcal{G}$ at leading non-trivial order, which in turn entails a complete calculation of the $1\rightarrow 3$ collinear splitting function in the presence of a colored medium. While this is already challenging when the medium is treated as a background color field \cite{Arnold:2021pin,Arnold:2022epx,Arnold:2023qwi,Arnold:2024whj}, the problem becomes even more complex here, as one would also need to include the medium particles that can affect the jet clustering. Consequently, we will rely on event-generator studies to address these questions. Before turning to simulations, however, we first build qualitative intuition using a simplified analytical model for energy-correlator measurements inside anti-$k_T$ jets.

\section{A simple model for edge effects}
\label{sec:analytical}

The goal of this section is to propose a simple functional form for the corrections to~\eqref{eq:refactorised} arising from edge effects. The resulting ansatz will later help  interpreting the event-generator results presented in Section~\ref{sec:MC}. To this end, we consider a simplified model of an anti-$k_T$ jet, idealized as circular around the $E$-scheme jet axis with a fixed jet radius $R$. Under this assumption, we analyze the semi-classical (leading-log) behavior of $\mathcal{G}$, while remaining agnostic to the specific collision system (p-p or A-A).

Although this section presents a simplified model, the discussion is more technical than the rest of the paper. Therefore, we summarize  the main result here so that readers may skip ahead to the results in Section~\ref{sec:MC} if desired. The key outcome of this section is the introduction of the following phenomenologically-motivated correction to the EEC spectrum in Eq.~\eqref{eq:refactorised} that accounts for edge effects. Specifically, in the range $R_L<R$,~\eqref{eq:refactorised} should be modified by a simple multiplicative factor
\begin{align}
    E(\langle \phi \rangle, R_L/R )= \left(1 - \langle \phi \rangle(p_T) \, K(R_L / R) \right) \,,
    \label{eq:edge_factor}
\end{align}
where $\langle \phi \rangle$ denotes the average deflection between the winner-takes-all axis and the $E$-scheme axis for the given sample of jets. Hence, $\langle \phi \rangle$ corresponds to the first moment of the axis-decorrelation distribution,   sometimes referred to as $\Delta j$ in the experimental literature \cite{ALICE:2022rdg,ALICE:2023dwg,CMS:2025dnx,CMS:2024msa}. This quantity depends on the jet clustering  procedure and, most importantly, on the jet transverse momentum, $p_T$. The function $K$ depends only on even powers of $R_L / R$, and is therefore independent of $p_T$. Both $\langle \phi \rangle$ and $K$ may also depend on the collision system considered. This ansatz effectively reduces, for a given collisions system, the unknown power corrections in~\eqref{eq:refactorised} to two single-variable functions, which can be determined either phenomenologically or through data-driven methods.

We now proceed with the discussion supporting this prescription. At leading log, we can replace the hard function by
\begin{align}
   H\left( x , \frac{p_T'^2}{\mu^2} \right) &\mapsto H\left( \frac{p_T'^2}{\mu^2} \right) \delta(x-1)\,,
\end{align}
so that \eqref{eq:factorization2} becomes
\begin{align} 
\frac{\td \sigma^{\rm LL}_{\rm EEC} (R)}{\td p_T \td R_L } = H ( p_T^2) ~ \mathcal{G} \left(p_T, R_L , R \right) \, ,
\label{eq:factorization2_LL}
\end{align}
where the dependence on the scale $\mu$ and the flavor indices have been omitted for simplicity. In the limit $R_L \ll R$, where $\mathcal{G}$ can be decomposed in terms of $\mathcal{J}$ and $j$, this leading-log expression reduces to~\eqref{eq:factorization_LL}.

The combined jet function $\mathcal{G}$ encodes the energy-weighted correlations between hadrons pairs ($i,k$) within a jet and can be expressed as a sum over contributions  from each pair:
\begin{align}
    \mathcal{G} = \sum_{i,k} \mathcal{G}_{ik}\,,
\end{align}
where $\mathcal{G}_{ik}$ denotes the contribution from the hadron pair ($i,k$). We decompose this sum into two parts:  one in which either $i$ or $k$ corresponds to the most energetic hadron, known as  ``winner-takes-all'' hadron, and another one containing  the remaining pairs. Labeling the hadrons by their $p_T$ such that the WTA hadron is labeled as $1$, we then write
\begin{align}
    \mathcal{G} = \mathcal{G}^{\rm WTA} + \sum_{i,k >1} \mathcal{G}_{ik}\,,
    \label{eq:Gdecomposed}
\end{align}
where the WTA-tagged combined jet function $\mathcal{G}^{\rm WTA}$ is defined as
\begin{align}
    \mathcal{G}^{\rm WTA} = 2 \sum_{k> 1} \mathcal{G}_{1 k}  + \mathcal{G}_{1 1}\,.
\end{align}
We will initially focus on $\mathcal{G}^{\rm WTA}$, developing an approach that then can be extended to the second term in \eqref{eq:Gdecomposed}. To proceed, we introduce two angular variables. The first,  $\phi$, is the angle between the WTA axis and the $E$-scheme jet axis, defined as
\begin{align}
    \phi \equiv \cos^{-1}(1-n_1\cdot n_{\rm jet})\,,
\end{align}
where $n_{1}$ and $n_{\rm jet}$ are unit light-like vectors along the  WTA axis and $E$-scheme jet axis, respectively. The second one, $\alpha_k$, is the angle between the $E$-scheme jet axis and the direction of the hadron $k$ being correlated with the WTA hadron, defined as
\begin{align}
    \alpha_k \equiv \cos^{-1}(1-n_k\cdot n_{\rm jet})\,,
\end{align}
where $n_{k}$ is a unit light-like vector along the direction of the hadron $k$. 

Within the circular jet approximation, $\mathcal{G}^{\rm WTA}$ can be computed semi-classically as
\begin{align}
    & \mathcal{G}^{\rm WTA}(p_T,R_L ,R) = \sum_{k} \int_0^R \! \! \td \phi \int_0^R \! \! \td \alpha_k \label{eq:GWTA}\\ 
    & ~~~~~~ \times \int^{1}_{0} \! \frac{\td z_1^2 \, \td z_k^2}{4} ~ P(z_1, z_k, y, R_L , \phi, \alpha_k | p_T)\,, \nonumber
\end{align}
where $P$ denotes the probability density for a jet of fixed $p_T$ to have an $E$-scheme jet axis $n_{\rm jet}$, a  WTA hadron with momentum $p_1 = p_T z_1 n_1$, and for an additional hadron to be observed with momentum $p_k = p_T z_k n_k$. In terms of cross-sections, $P$ is given by
\begin{align}
    P  \equiv \left(\frac{\td \sigma}{\td p_T }\right)^{-1} \frac{\td \sigma}{\td z_1 \, \td z_k \, \td R_L \, \td \phi \, \td \alpha_k \, \td p_T}\,,
\end{align}
where $\sigma$ is the inclusive cross-section to produce hadrons $1$ and $k$ plus any additional hadrons $X$.

The probability distribution $P$ can be decomposed into two parts: one describing the deflection of the WTA axis relative to the $E$-scheme jet axis, and a conditional probability describing the configuration of the hadrons given this deflection:
\begin{align}
    P = \left(\frac{\td \sigma}{\td p_T }\right)^{-1} \frac{\td \sigma }{ \td \phi \, \td p_T} ~ \tilde{P}(z_1, z_k, R_L , \alpha_k | \phi, p_T) \,,
\end{align}
where $\td \sigma /(\td \phi \, \td p_T)$ gives the probability for the WTA axis to be deflected by an angle $\phi$ with respect to the $E$-scheme jet axis. This cross-section is sharply-peaked near $\phi = 0$, since the WTA axis is typically closely aligned with the jet axis \cite{Cal:2019gxa,ALICE:2022rdg,ALICE:2023dwg,CMS:2025dnx,CMS:2024msa}. The conditional distribution $\tilde{P}(z_1, z_k, R_L , \alpha_k | \phi, p_T)$ describes, for a jet with fixed $p_T$, the probability of observing  a WTA hadron with energy fraction $z_1$ and an additional hadron with energy fraction $z_k$, where the two hadrons are separated by an angle $R_L$ and the additional hadron lies at an angle $\alpha_k$ from the jet axis, given that the WTA axis is deflected by an angle $\phi$ from the $E$-scheme jet axis.  This distribution must satisfy the standard conditional probability normalization, 
\begin{align}
    \int^{1}_{0} \! \! \td z_1 \, \td z_2 \int^{\pi}_{0} \td R_L \! \! \int^{\pi}_{0} \! \! \td \alpha_k \, \tilde{P}(z_1, z_k, R_L , \alpha_k | \phi, p_T) = 1 \,,
\end{align}
ensuring that the total probability over all allowed hadron configurations sums to unity.

The cross-section $\td \sigma /(\td \phi \, \td p_T)$ is sharply peaked near $\phi=0$, and therefore can be expanded in moments as
\begin{align}
    \left(\frac{\td \sigma}{\td p_T }\right)^{-1} \frac{\td \sigma }{ \td \phi \, \td p_T} & = \sum_{n} \frac{1}{n!} (-1)^n \langle \phi^n \rangle (p_T) ~ \delta^{(n)}(\phi)\,,
\end{align}
where $\delta^{(n)}$ is the $n$-th derivative of the Dirac delta function, and 
\begin{align}
\langle \phi^n \rangle (p_T) = \, \int \td \phi \, \phi^n \, \left(\frac{\td \sigma}{\td p_T}\right)^{-1} \frac{\td \sigma}{\td \phi \, \td p_T} 
\end{align}
is the $n$-th moment of the WTA axis to jet axis deflection distribution $\td \sigma /(\td \phi \, \td p_T)$  for a jet with transverse momentum $p_T$. The leading term ($n$=0) corresponds to the configuration where the WTA and $E$-scheme axis are aligned, while higher moments encode the effect of deflections between the two axes. Since the distribution is sharply peaked near $\phi=0$, there is a hierarchy in the moments $\langle \phi^n \rangle \ll \langle \phi^{n-1} \rangle$. We note that the distribution for the jet axis decorrelation $\td \sigma /(\td p_T \td \phi)$ has been measured  and is sometimes referred to as $\Delta j$ \cite{ALICE:2022rdg,ALICE:2023dwg,CMS:2025dnx,CMS:2024msa}. The moments $\langle \phi^n \rangle $ are simply moments of those measurements.

Introducing this moment expansion into~\eqref{eq:GWTA} allows us to perform the bounded  integration over $\phi$, yielding
\begin{align}
    &\mathcal{G}^{\rm WTA}(p_T,R_L ,R) = \nonumber \\ 
    & \sum_k \! \int_0^R \! \! \td \alpha_k \int \! \frac{\td z_1^2 \, \td z_k^2}{4} \, \tilde{P}(z_1, z_k, R_L , \alpha_k | 0, p_T)\, \delta(\alpha_k - R_L ) \nonumber \\
    & - \langle \phi \rangle (p_T) \sum_k \! \int_0^R \! \! \td \phi ~ \delta^{(1)}(\phi) \int_0^R \! \! \td \alpha_k \int \! \frac{\td z_1^2 \, \td z_k^2}{4} \,  \tilde{P}  \nonumber \\  &\,+\mathcal{O}\left(\langle \phi^2 \rangle \right),
    \label{eq:GWTA_2} 
\end{align}
where the first term corresponds to the computation of the EEC in the limit where the jet axis coincides with the WTA axis, independently of the measured correlations. This limit is given by single highly-energetic parton propagating  along the jet axis, which then fragments collinearly into the WTA hadron and another hadron whose correlation is measured. This configuration therefore captures the leading-logarithmic limit of the WTA-tagged EEC at small $R_L$, i.e., the case where all measured correlations involve the WTA hadron. Consequently, this term factorizes with the same functional form as given in~\eqref{eq:factorization_LL}
\begin{align}
&\sum_k \! \int_0^R \! \! \td \alpha_k \int \! \frac{\td z_1^2 \, \td z_k^2}{4} \, \tilde{P}(z_1, z_k, R_L , \alpha_k | 0, p_T)\, \delta(\alpha_k - R_L ) \nonumber \\
&\approx
\mathcal{J} \left(\ln \! \frac{ R^2 p_T^2}{\mu^2} \right)  \frac{j^{\rm WTA}\left(\ln \! \frac{ R_L^2 p_T^2}{\mu^2} \right)}{R_L } \Theta(R_L < R)\,, \label{eq:term1}
\end{align}
where $j^{\rm WTA}$  denotes the WTA-tagged EEC jet function, defined in analogy to $\mathcal{G}^{\rm WTA}$ in~\eqref{eq:Gdecomposed}, and the $\Theta$ function arises from satisfying the $\alpha_k$ integration limits.

Let us now look at the second term in \eqref{eq:GWTA_2}, first considering its limiting behavior. As $R_L \to 0$, the separation between the two hadrons becomes small, entering the di-hadron fragmentation regime \cite{Rogers:2024nhb,Pitonyak:2023gjx,Pitonyak:2025lin,Lee:2025okn,Chang:2025kgq,Kang:2025zto}. In this limit, the two hadrons are produced locally from a single decaying parton and factorize from the rest of the event. Consequently, as $R_L \rightarrow 0$ it is required that $\tilde{P}$ becomes independent of $\phi$ and the integration over $\phi$ in the second term of \eqref{eq:GWTA_2} is zero. This term is therefore only non-zero away from the $R_L \rightarrow 0$ limit and so represents a contribution to the $R_L/R$ power corrections we aim to understand. For unpolarized jet production, the power corrections to Eq.~\eqref{eq:refactorised} are expected to scale as $(R_L/R)^{2n}$ with $n\geq 1$ relative to the leading term  \cite{Kang:2016mcy,Lee:2024icn}. 

We assume that the dominant contribution of the edge effects can be captured by a multiplicative correction to the EEC  measurement. With this in mind, we write this term as
\begin{align}
&\sum_k \! \! \int_0^R \! \! \td \phi ~ \delta^{(1)}(\phi) \int_0^R \! \! \td \alpha_k \int \! \frac{\td z_1^2 \, \td z_k^2}{4} \,  \tilde{P}(z_1, z_k, R_L , \alpha_k | \phi, p_T) \nonumber \\
&\approx \mathcal{J} \left(\ln \frac{ R^2 p_T^2}{\mu^2} \right)  \frac{j^{\rm WTA}\left(\ln \frac{ R_L^2 p_T^2}{\mu^2} \right)}{R_L } \Theta(R_L < R) \, K\left(\frac{R_L}{R}\right)\, ,
\label{eq:term2}
\end{align}
where $K$ is an unknown function. 
Since~\eqref{eq:term2} must obey the same renormalization-group evolution as Eq.~\eqref{eq:term1}, 
factoring out the product $\mathcal{J} \,  j^{\rm WTA}$, requires $K$ to be independent of $p_T/\mu$. Consequently, $K$ can only depend on even powers of $R_L /R$ and encodes only the shape of the edge effects, while their magnitude as a function of the $p_T$ is determined by the average separation between the WTA and $E$-scheme axes $\langle \phi \rangle (p_T)$.

In summary, for $R_L < R$, we are led to the ansatz
\begin{align}
    &\mathcal{G}^{\rm WTA} \approx \nonumber \\
    &\mathcal{J} \left(\ln \frac{ R^2 p_T^2}{\mu^2} \right) \frac{j^{\rm WTA}\left(\ln \frac{ R_L^2 p_T^2}{\mu^2} \right)}{R_L } \left[1 - \langle \phi \rangle ( p_T ) ~ K (R_L /R) \right] \, ,
\end{align}
with the flavor indices still omitted.

We can now turn our attention to the remainder of the jet function. Having analyzed the $\mathcal{G}^{\rm WTA}$ term in \eqref{eq:Gdecomposed}, we can apply the same approach to $\mathcal{G}_{ik}$ where $i,k > 1$, i.e. to pairs where neither particle is the WTA hadron. We do this by decomposing the sum over $\mathcal{G}_{ik}$ into pairs involving the next-to-WTA hadron (NWTA) and the remaining pairs $i,k > 2$. Let the angle between the WTA hadron and the next-to-WTA hadron be $\delta$. By expanding in both $\phi$ and $\delta$, a simple generalization of the previous discussion leads to 
\begin{align}
    \mathcal{G}^{\rm NWTA} \approx & \,\mathcal{J} \left(\ln \frac{ R^2 p_T^2}{\mu^2} \right) \, \frac{j^{\rm NWTA} \left(\ln \frac{ R_L^2 p_T^2}{\mu^2} \right)}{R_L }\nonumber \\
    & \,\,\times \left[1 - (\langle \phi \rangle + \langle \delta \rangle) \,K(R_L /R)\right] \, ,
\end{align}
where we have suppressed the $p_T$-dependence on $\langle \phi \rangle$ and $\langle \delta \rangle$. 

This procedure can be iterated to the next-to-next-to WTA hadron, and so forth. Summing these contributions, we find 
\begin{align}
    &\mathcal{G}(p_T,R_L ,R) \approx \mathcal{J} \left(\ln \frac{ R^2 p_T^2}{\mu^2} \right) \frac{j\left(\ln \frac{ R_L^2 p_T^2}{\mu^2} \right)}{R_L } \nonumber  \\&\times\left[1 - \langle \phi \rangle \,  K(R_L /R ) + \mathcal{O}\left(\alpha_{\rm s}\langle \delta \rangle ,  \langle \phi^2 \rangle\right)\right]\, ,
    \label{eq:EEC_beyondLL}
\end{align}
where we note that the $\mathcal{G}^{\rm NWTA}$ is suppressed by $\alpha_{\rm s}$ with respect to $\mathcal{G}^{\rm WTA}$ \footnote{We note that, in the presence of a colored medium, additional partons could be non-perturbatively generated from the medium and therefore not suppressed by factors of $\alpha_{\rm s}$. Nevertheless, these contributions should still be suppressed in an EEC measurement, as such particles carry low energies.}. Since both $\langle \delta \rangle$ and $\alpha_{\rm s}$ are individually small, their product is negligible for the purposes of this work.

Hence, starting from a simple circular jet model and  treating the EEC factorization semi-classically, we are lead to a simple functional form  describing  edge effects in EEC measurements, which should be applicable to any collisions system. Going beyond the leading-log limit, this ansatz modifies Eq.~\eqref{eq:factorization2} as 
\begin{align}
    \frac{\td \sigma_{\rm EEC}}{\td R_L \td p_T} \approx & \,  H_a \otimes \mathcal{J}_{ab} \boxtimes 
     \frac{j_b}{R_L } \left[1  - 
       \langle \phi \rangle \, K(R_L /R) \right]\,, 
       \label{eq:final}
\end{align}
where $\otimes$ denotes the convolution over $x$ in \eqref{eq:refactorised}, $\boxtimes$ the convolution over $y$, and we have suppressed the standard dependencies of $\mathcal{J}$, $j$ and $\langle \phi \rangle$. 
While the structure of Eq.~\eqref{eq:final} applies to any collision system, the functions $K$ and $\langle \phi \rangle$ may depend on the collision system considered. We will assign labels to both to make the collision system clear, e.g. $K^{\rm AA}$ and $\langle \phi \rangle^{\rm AA}$.

We can now consider the ratio between two equivalent EEC measurements, one performed in A-A and the other in p-p collisions. Using \eqref{eq:final}, we find that
\begin{align}
    \frac{\td \sigma^{\rm AA}_{\rm EEC}/\td R_L }{\td \sigma^{\rm pp}_{\rm EEC}/\td R_L } \approx & \frac{ H_a^{\rm AA} \otimes \mathcal{J}_{ab}^{\rm AA} \boxtimes 
     j_b^{\rm AA} }{H_a^{\rm pp} \otimes \mathcal{J}_{ab}^{\rm pp} \boxtimes 
     j_b^{\rm pp}} \nonumber \\
     & \times \left[1 - \left(\langle\phi \rangle^{\rm AA} K^{\rm AA} - \langle \phi \rangle^{\rm pp} K^{\rm pp}\right) \right]\, .\label{eq:doubleratio}
\end{align}
It is immediately notable that edge effects are suppressed in the ratio, since they enter only through the combination $\langle\phi\rangle K$, which is positive in any collision system. Furthermore, if $\langle\phi \rangle K$ is relatively independent of the process that initiates the jet, edge effects are expected to largely cancel when taking the A-A to p-p ratio. Indeed, while $\td \sigma / (\td \phi \, \td p_T)$ varies significantly with the collision system, experimental measurements of this distribution in Pb-Pb and p-p  \cite{ALICE:2022rdg,ALICE:2023dwg,CMS:2025dnx,CMS:2024msa} show that its first moment, $\langle\phi \rangle$, is largely independent of the system. For this reason, our event generator studies  will focus first on verifying this ansatz and then on  understanding the behavior of $K(R_{L}/R)$.

This result motivates the study of the following key questions using  event generators. First, do edge effects in the jet EEC spectrum scale linearly with the average deflection between the WTA axis and $E$-scheme axis across different collision systems and modeling choices? Second, does $K$ depend only on even powers of $R_L /R$ in all collision systems? Affirmative answers to these questions, as we will  obtain in the upcoming section, support the ansatz introduced above. Third, how does $K$ vary with the initiating process, or equivalently, how does the shape of edge edge effects in the EEC vary across simulations with different event generators, and by how much are these effects reduced when taking the A-A to p-p ratio? Finally, can any residual edge effects in heavy-ion measurements be controlled or mitigated through a simple understanding of $K$? Addressing these questions allows us to quantify the theoretical uncertainty in EEC calculations of the form~\eqref{eq:refactorised}, arising from neglected edge effects, ensuring that such omissions do not contaminate extractions of medium properties or matrix elements when comparing to experimental data. Furthermore, in heavy-ion generator-data comparisons, examining  $\td \sigma / (\td \phi \, \td p_T)$ in the generator against available measurements  \cite{ALICE:2022rdg,ALICE:2023dwg,CMS:2025dnx,CMS:2024msa} provides a practical tool to validate the implementation of edge effects, thereby enabling a more robust interpretation of jet substructure measurements.

%%%%%%%%%%%%%%%%%%%%%%%%%%%%%%%%%
%%%%%%%%%%% MC section %%%%%%%%%
\section{Monte Carlo results}
\label{sec:MC}

In this section, we analyze edge effects in the two-point energy correlator for both p-p and Pb-Pb collisions using event generator simulations.
For p-p, we use inclusive jet samples from $\sqrt{s}=5.02$ TeV p-p collisions generated with \textsc{Pythia} 8.230  \cite{Sjostrand:2014zea} and \textsc{Herwig}  7.2.2 \cite{Bellm:2015jjp,Bahr:2008pv} Monte Carlo event generators.   The \textsc{Pythia}8 samples consist of around 44 million events generated using the CP5 tune \cite{CMS:2019csb}, while  \textsc{Herwig}7 samples include about 7 million events  were generated with tune CH3 \cite{CMS:2020dqt}. Jets are clustered from final-state hadrons using the anti-$k_T$ algorithm \cite{Cacciari:2008gp}, and the EEC is computed using charged particles with $p_T^{\rm ch} > 1$ GeV.

\begin{figure}
\includegraphics[scale=0.47]{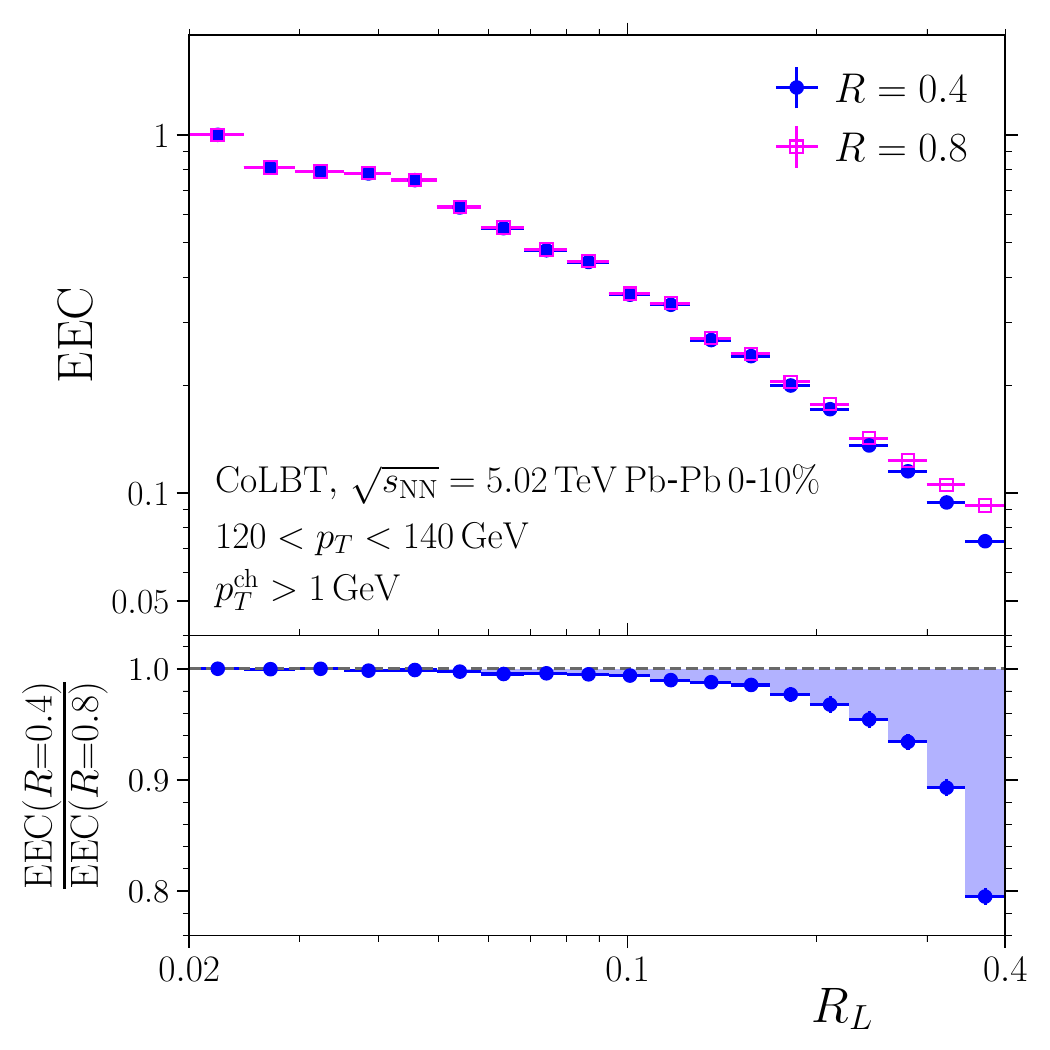}
\caption{Top: Two-point energy correlator for anti-$k_T$ jets with $120 < p_T < 140$~GeV, computed using the same sample of events with the \textsc{CoLBT} in $\sqrt{s_{\rm NN}} = 5.02$ TeV  0-10$\%$ Pb-Pb collisions, for jet radii $R = 0.4$ (blue circles), and $R = 0.8$ (pink squares). Bottom: ratio between the EEC for $R = 0.4$ and $R = 0.8$.}
\label{fig2}
\end{figure}

The Pb-Pb analysis uses  inclusive samples from $\sqrt{s_{\rm NN}}=$ 5.02 TeV Pb-Pb collisions  generated with the Coupled Linear Boltzmann Transport hydrodynamic (CoLBT)  model \cite{Chen:2017zte,Chen:2020tbl, Zhao:2021vmu}  and with \textsc{JEWEL}~\cite{Zapp:2008gi,Zapp:2012ak,Zapp:2013vla,KunnawalkamElayavalli:2017hxo}. The \textsc{CoLBT} event generator combines a \textsc{Pythia}8 vacuum shower (Monash tune) with the LBT  quenching model \cite{He:2015pra,Luo:2023nsi} embedded within the (3+1)D viscous hydrodynamics CLVisc  \cite{Pang:2018zzo}. This setup allows for a concurrent simulation of the jet propagation and modification within the QGP, as well as the medium's event-by-event evolution and its back-reaction to the jet, known as medium response. All CoLBT results presented in this manuscript correspond to the  for the 0-10$\%$ centrality class of  $\sqrt{s_{\rm NN}}=$ 5.02 TeV Pb-Pb collisions. \textsc{JEWEL} is a Monte Carlo event generator that models perturbatively the evolution and interactions of QCD jets with a thermal medium, which is created with a spacial profile based on the nuclear overlap function. We utilize v$2.4$ of \textsc{JEWEL} with an initial temperature $T_i = 0.5$ GeV and a centrality corresponding to 0-10$\%$. The results presented in the main text correspond to the \textsc{JEWEL} configuration including medium recoils, while Appendix~\ref{appB} shows the corresponding results for the configuration without recoils. One additional note is that \textsc{JEWEL} includes an option to subtract the thermal component of the recoil partons that are produced during the jet-medium interactions. Since we are studying the ratios of the correlators at fixed $R_L$, we did not apply this subtraction and instead keep this contribution within the two radius selections. This choice effectively leaves to an overestimation of the impact of the medium recoils, as they do not further re-scatter and also include their pre-interaction energy of the thermal partons.
As in the p-p case, for both heavy-ion event generators jets are clustered from final-state hadrons using the anti-$k_T$ algorithm and the EEC is computed using charged particles with $p_T^{\rm ch} > 1$ GeV.

Before presenting the full analysis, we first illustrate edge effects in heavy-ion collisions in a manner analogous to what was done for p-p collisions in the introduction. Chidi constructs the EEC on a sample of anti-$k_T$ jets with radius $R=0.4$, generated with the \textsc{CoLBT} event generator for  0-10$\%$ central Pb-Pb collisions at $\sqrt{s_{\rm NN}} = 5.02$ TeV. Since edge effects arise from the jet reconstruction procedure itself, they should affect Chidi's \textsc{CoLBT} analysis as well. To investigate this, Eleanor repeats the analysis on the same sample of events but with a larger radius $R=0.8$. Chidi’s $R=0.4$ result (blue circles) and Eleanor’s $R=0.8$ result (pink squares) are shown in the top panel of Fig.~\ref{fig2}, with their differences arising only from edge effects. As in the p-p case shown in Fig.~\ref{fig1}, deviations from unity in the EEC ratio, highlighted with the blue shaded area, arise from correlations missed by Chidi’s $R=0.4$ analysis, which are recovered in Eleanor’s $R=0.8$ measurement. 

\begin{figure}
\includegraphics[scale=0.47]{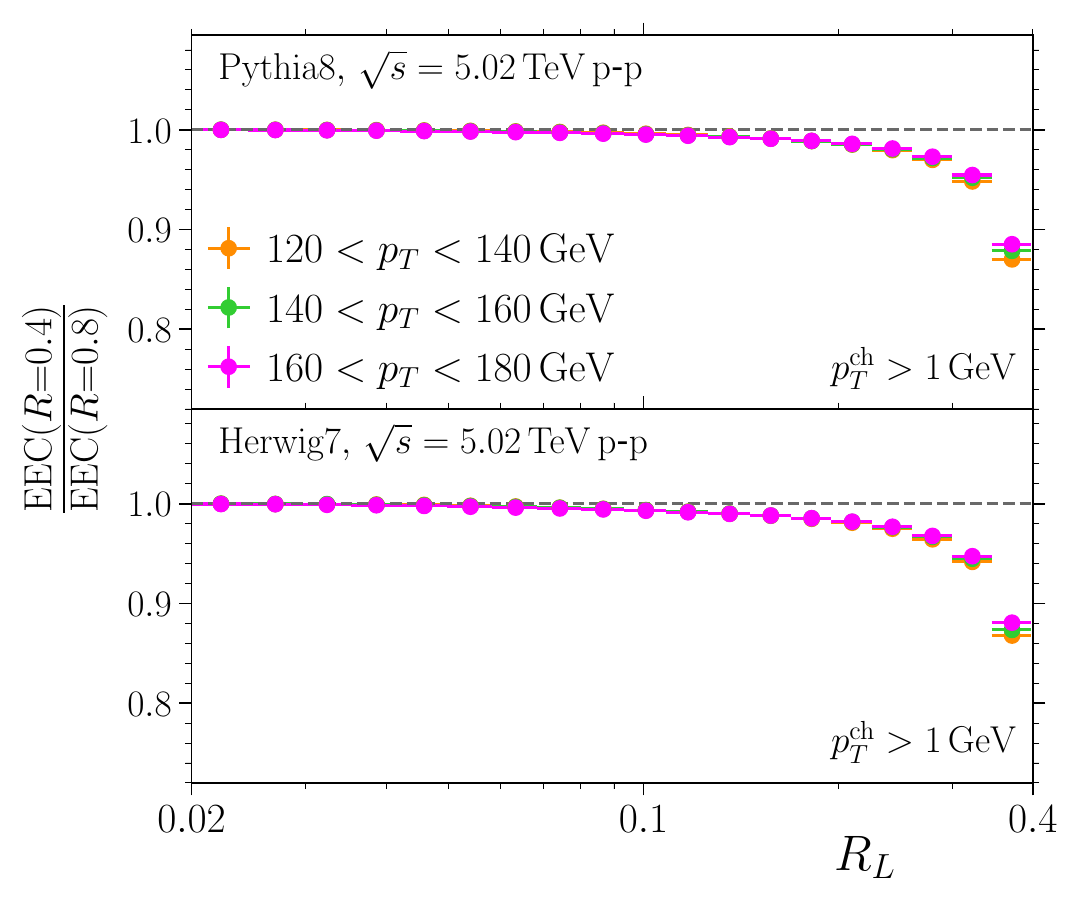}
\caption{Top: ratio of the EEC computed within anti-$k_T$ jets of radius $R = 0.4$ to that obtained with $R = 0.8$,  using the same sample of events generated with \textsc{Pythia}8 in $\sqrt{s} = 5.02$ TeV p-p collisions, shown for three different jet $p_T$ bins. Bottom: same as top, but using \textsc{Herwig}7.}
\label{fig:EEC_ratio_pp}
\end{figure}
We now quantify the edge effects impacting Chidi's p-p and Pb-Pb measurements by analyzing the ratio of the EEC for jets  with radius $R = 0.4$ to those with $R = 0.8$, using different even generators in both p-p and Pb-Pb collisions. We begin with the p-p results. Fig.~\ref{fig:EEC_ratio_pp} shows this ratio for three jet $p_T$ intervals, with jets  generated with \textsc{Pythia}8 (top panel) and \textsc{Herwig}7  (bottom panel) Monte Carlo event generators.  We stress that, for a given event generator and  $p_T$ interval, the same sample of events is used for both radii, so deviations from unity in the EEC($R = 0.4$)/EEC($R = 0.8$) ratio at large $R_L$ reflect only edge effects.  As expected, edge effects grow at larger angular separations and lower jet $p_T$, with only minor differences observed between \textsc{Pythia}8 and \textsc{Herwig}7 in this figure. 
\begin{figure}
\includegraphics[scale=0.47]{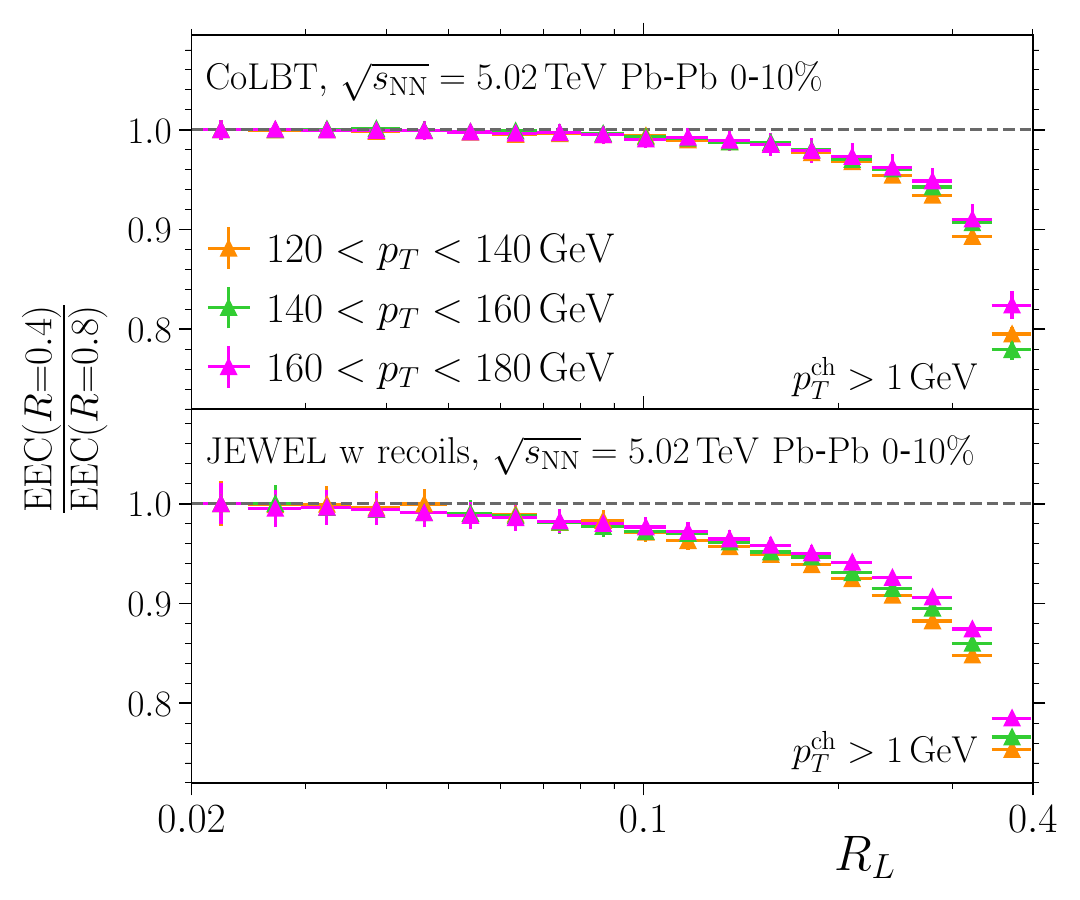}
\caption{Top: ratio of the EEC computed within jets with radius $R = 0.4$ to that obtained with $R = 0.8$,  using the same sample of jets generated with the \textsc{CoLBT} in $\sqrt{s_{\rm NN}} = 5.02$ TeV  0-10$\%$ Pb-Pb collisions, shown for three different jet $p_T$ bins. Bottom: same as top, but using \textsc{JEWEL} with recoils.}
\label{fig:EEC_ratio_PbPb}
\end{figure}

We now turn to Fig.~\ref{fig:EEC_ratio_PbPb}, which  shows the same ratio in Pb-Pb collisions. The results are  obtained for inclusive samples in  $\sqrt{s_{\rm NN}}=$ 5.02 TeV  0-10$\%$ central Pb-Pb collisions  generated with the \textsc{CoLBT}  model (top panel) and with \textsc{JEWEL} with recoils (bottom panel). As in the p-p case, deviations from unity in the Pb-Pb EEC($R=0.4$)/EEC($R=0.8$) ratio shown in Fig.~\ref{fig:EEC_ratio_PbPb} arise from correlations missed by the smaller jet radius and recovered when increasing the radius to $R=0.8$. These deviations become more pronounced at lower jet $p_T$ and, for a given $p_T$ bin, are found to be larger in \textsc{JEWEL} (with recoils) than in CoLBT.  Furthermore, comparing to Fig.~\ref{fig:EEC_ratio_pp}, edge effects, for a given $p_T$ bin, are larger in the heavy-ion event generators than in the p-p ones.

To systematically compare the magnitude of these effects across generators, we define a differential measure quantifying the total deviation of the ratio ${\rm EEC}(R=0.4)/{\rm EEC}(R=0.8)$ from unity for a given $R_L$ bin and jet $p_T$ bin, as
\begin{equation}
\Delta_{\rm edge\,effects} (R_L) =  1 - \frac{\td \sigma^{R=0.4}_{\rm EEC}/\td R_L }{\td \sigma^{R=0.8}_{\rm EEC}/\td R_L }\,.
\label{eq:edge_effects}
\end{equation}
Substituting the ansatz in~\eqref{eq:final}, we aim to test, for both  Pb-Pb and p-p, the validity of the relation
\begin{align}
\Delta_{\rm edge\,effects} (R_L) & \approx  \langle\phi \rangle^{R=0.4} K\left(\frac{R_{L}}{0.4} \right) - \langle \phi \rangle^{R=0.8} K\left(\frac{R_{L}}{0.8} \right) ,
\nonumber \\
 & \approx \langle\phi \rangle \, K\left(\frac{R_{L}}{0.4} \right) = 1 - E\big( \langle\phi \rangle , R_{L}/0.4 \big)\,,
\label{eq:expectation}
\end{align}
where in the second line we 
dropped the jet-radius label on $\langle \phi \rangle$ since it is nearly  independent of $R$ in the regime $\langle \phi \rangle \ll R$, and we used   Eq.~\eqref{eq:edge_factor}. 
In the following, we restrict our analysis to the five highest $R_L$ bins in Figs.~\ref{fig:EEC_ratio_pp} and~\ref{fig:EEC_ratio_PbPb}, where edge effects are most significant.

\begin{figure}[t]
\includegraphics[scale=0.47]{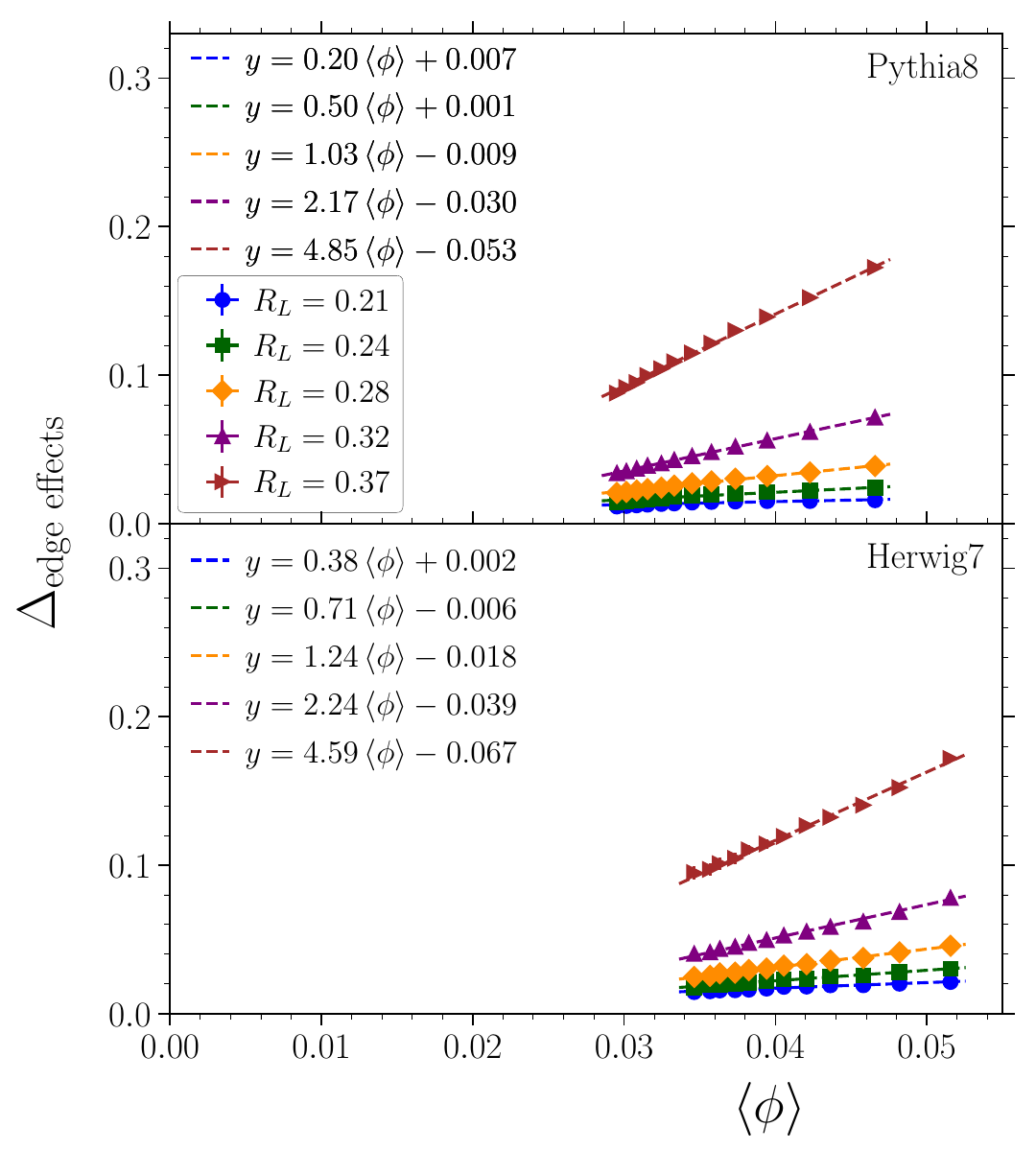}
\caption{Top: edge effects defined as in \eqref{eq:edge_effects} for the highest five $R_L$ bins shown as a function of the average $E$-scheme WTA-axes difference  $\langle \phi \rangle$  for \textsc{Pythia}8  in $\sqrt{s} = 5.02$ TeV p-p collisions. Bottom: same as top but for \textsc{Herwig}7. Linear fits to both \textsc{Pythia}8 and \textsc{Herwig}7 are included. The corresponding fit parameters, including their errors, are listed in Table~\ref{tab:fits_binbybin_pp} of Appendix~\ref{appA}.}
\label{fig:Edge_deltaphi_last5bins_pp}
\end{figure}

For each of these five $R_L$ bins in Fig.~\ref{fig:EEC_ratio_pp}, we compute $\Delta_{\rm edge\,effects}$ for each $p_T$ bin considered. Additionally, for a given $p_T$ bin, we determine the average angular difference $\langle \phi \rangle$ between the WTA and $E$-scheme axes. We then plot  $\Delta_{\rm edge\,effects}$ as a function of $\langle \phi \rangle$ for each $R_L$ bin. These results are shown for \textsc{Pythia}8 (Herwig7) in the top (bottom) panel of Fig.~\ref{fig:Edge_deltaphi_last5bins_pp}, where the five colored markers represent the five $R_L$ bins.  Although Fig.~\ref{fig:EEC_ratio_pp} showed only three $p_T$ bins  for clarity, Fig.~\ref{fig:Edge_deltaphi_last5bins_pp} includes additional bins. Each point along the $\langle \phi \rangle$ axis corresponds to one of twelve $p_T$ bins considered, covering the $60 < p_T < 300$ GeV regime in 20 GeV intervals. Since $\langle \phi \rangle$ increases as the jet $p_T$ decreases,  the largest  $\langle \phi \rangle$ point in this figure correspond to the $60 < p_T < 80$ bin. For each $R_L$ bin, we also show linear fits, indicated by dashed lines matching the corresponding bin color, of the form
\begin{equation}
\Delta_{\rm edge\,effects}\,(R_L) = K_0(R_L) + K(R_L) \, \langle \phi \rangle\, .
\label{eq:linear_fit}
\end{equation}
Validating Eq.~\eqref{eq:expectation} requires that the extracted $K_0(R_L)$ is consistent with zero. The fitted parameters $K_0$ and $K$ for each $R_L$ bin are listed in the legend of the top-left figure panels, with uncertainties provided in Appendix~\ref{appA}.

\begin{figure}[t]
\includegraphics[scale=0.47]{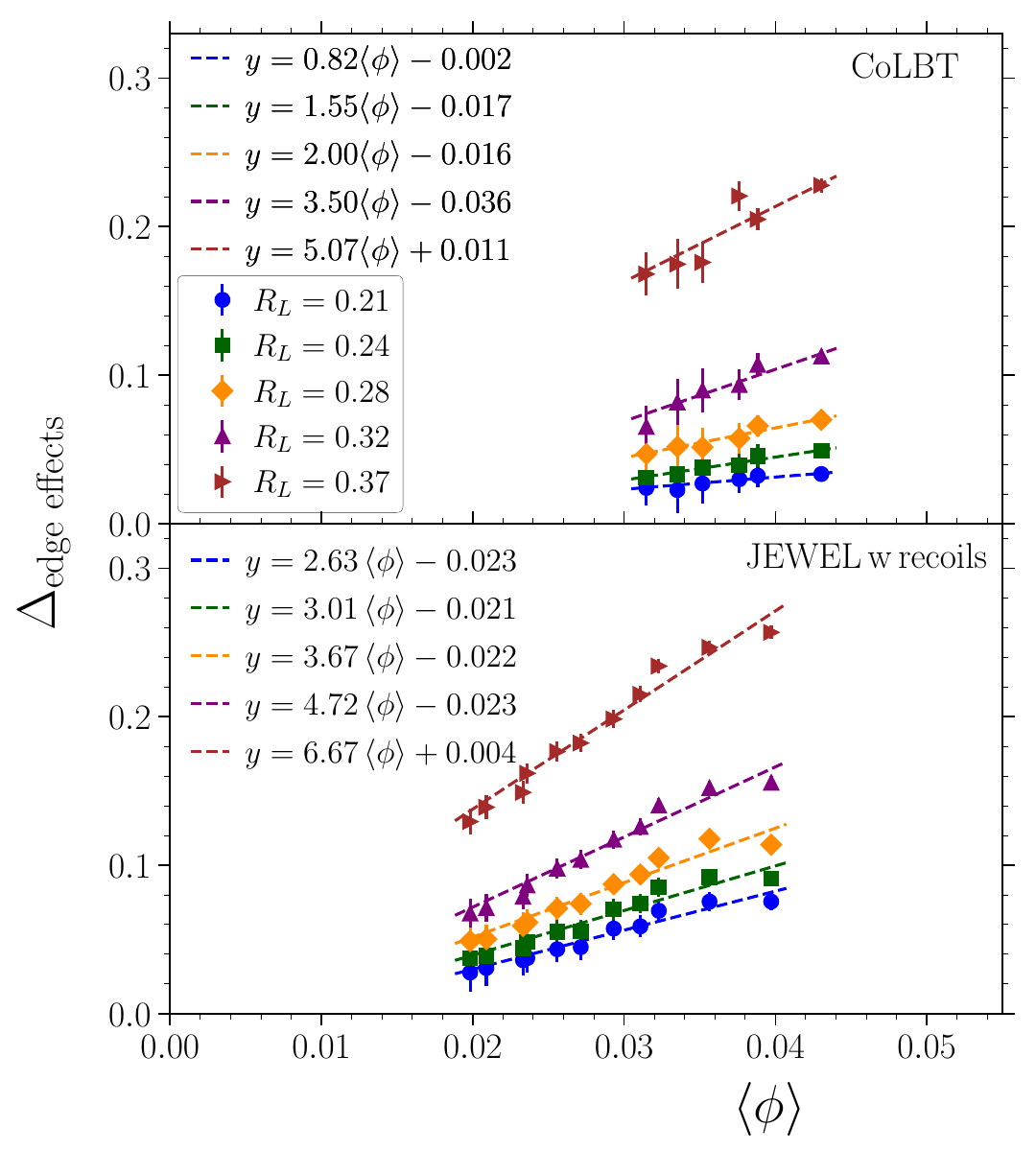}
\caption{Top: edge effects defined as in \eqref{eq:edge_effects} for the highest five $R_L$ bins shown as a function of the average $E$-scheme WTA axes difference  $\langle \phi \rangle$  for \textsc{CoLBT} in $\sqrt{s_{\rm NN}} = 5.02$ TeV  0-10$\%$ Pb-Pb collisions. Bottom: same as top but for \textsc{JEWEL} with recoils. Linear fits to both \textsc{CoLBT} and \textsc{JEWEL} are included. The corresponding fit parameters, including their errors, are listed in Table~\ref{tab:fits_binbybin_PbPb} of Appendix~\ref{appA}.}
\label{fig:Edge_deltaphi_last5bins_PbPb}
\end{figure}

A few comments comparing Fig.~\ref{fig:Edge_deltaphi_last5bins_pp} and Eq.~\eqref{eq:expectation} are in order. 
The linear fits confirm the expected linear dependence, with their resulting positive slopes indicating that edge effects suppress the EEC spectrum. As shown in the figure legend, the fitted $K$-values grow with increasing $R_L$, reflecting the stronger impact of edge effects near the jet boundary. Moreover, for a fixed $R_L$, the slopes obtained with \textsc{Herwig}7 and \textsc{Pythia}8 are similar but not identical, reflecting modeling-dependent differences associated with differences in the shower, the hadronization and tune choices. Finally, as anticipated from Eq.~\eqref{eq:expectation}, the fitted intercepts are all close to zero or compatible with it within uncertainties (see Table~\ref{tab:fits_binbybin_pp} of Appendix~\ref{appA}), providing further validation of~\eqref{eq:expectation}.

We now turn to the heavy-ion results  presented in Fig.~\ref{fig:Edge_deltaphi_last5bins_PbPb}, which is analogous to Fig.~\ref{fig:Edge_deltaphi_last5bins_pp}, but for 0-10$\%$ central Pb-Pb collisions at $\sqrt{s_{\rm NN}} = 5.02$ TeV simulated with \textsc{CoLBT} (top) and \textsc{JEWEL} with recoils (bottom). In both cases, the available statistics are smaller than in the p-p MC, resulting in fewer  $p_T$ bins, as reflected from the reduced number of data points per $R_L$ bin. For \textsc{JEWEL} (bottom panel), eleven  $p_T$ bins were analyzed, spanning the $100 < p_T < 320~$GeV regime in $20~$GeV intervals. \textsc{CoLBT} statistics are further limited due to the computational complexity of the model, which includes a full treatment of medium response via a concurrent simulation of the hard partons and hydrodynamic evolution. As a result, only six $p_T$ intervals are considered: $100 < p_T < 120$, $120 < p_T < 140$, $140 < p_T < 160$, $160 < p_T < 180$, $180 < p_T < 200$, and $200 < p_T < 260$~GeV, with the last bin being wider due to the reduced statistics at higher $p_T$.

As in the p-p case, the MC Pb-Pb  results in Fig.~\ref{fig:Edge_deltaphi_last5bins_PbPb} are fitted using Eq.~\eqref{eq:linear_fit}, with the extracted fit parameters shown in the figure legends (see Table~\ref{tab:fits_binbybin_PbPb} of Appendix~\ref{appA} for their errors). The fits indicate that edge effects increase linearly with the average separation between the WTA and $E$-scheme axes, with intercepts close to or compatible with zero, further validating~\eqref{eq:expectation} across collision systems. For a given $R_L$ bin, the fitted  $K$ values  differ notably between \textsc{JEWEL} with recoils and CoLBT, with \textsc{JEWEL} exhibiting larger slopes, indicative of stronger edge effects. Overall, the edge effects in the Pb-Pb MC considered here are more pronounced than in p-p.

\begin{figure}
\includegraphics[scale=0.47]{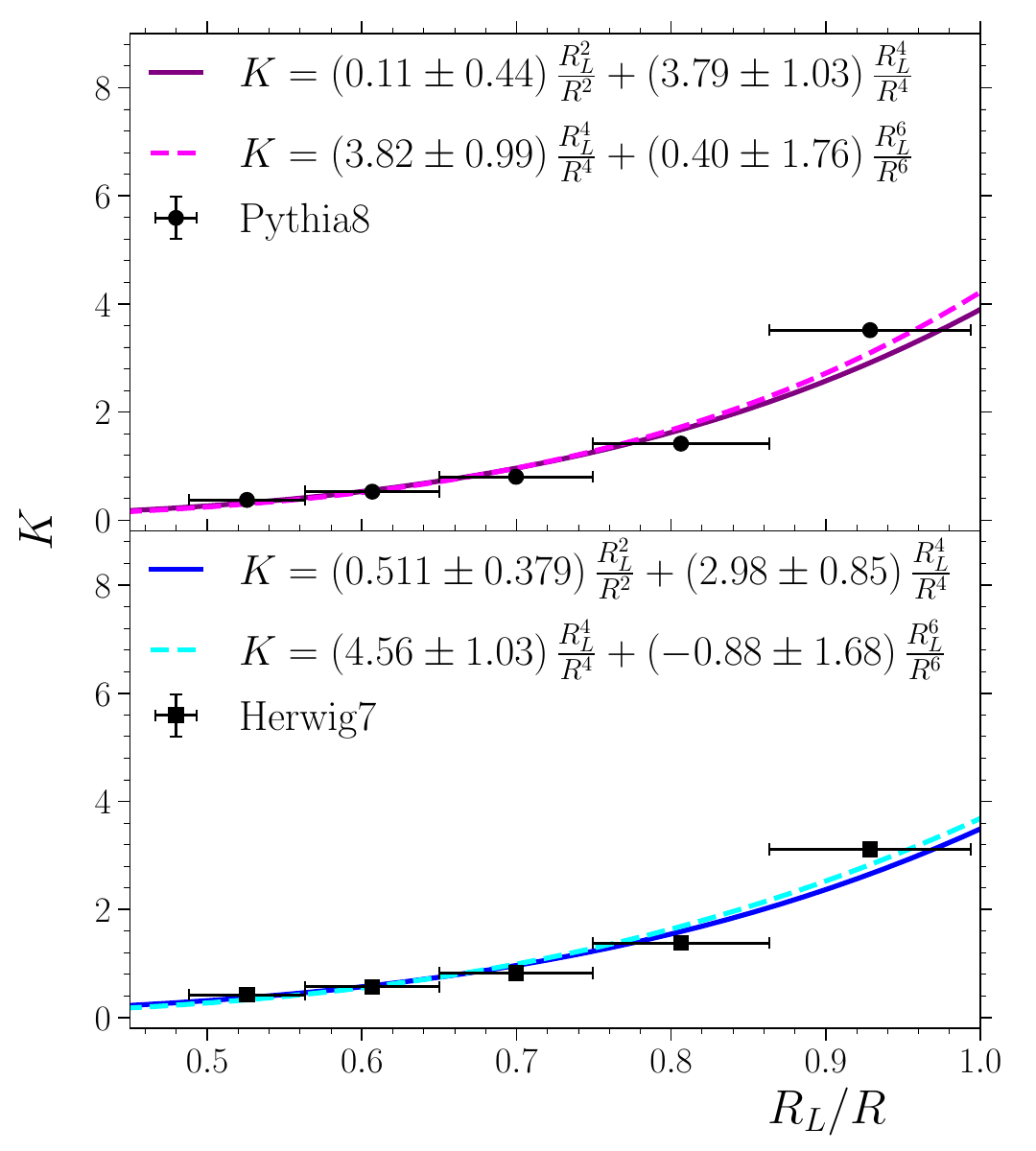}
\caption{Top: fitted $K$-value  for the highest five $R_L$ bins as a function of $R_L/R$ with $R=0.4$  for \textsc{Pythia}8  in $\sqrt{s} = 5.02$ TeV p-p collisions. Bottom: Same as top but for \textsc{Herwig}7. Solid lines show the  fits following~\eqref{eq:quartic_fit} for \textsc{Pythia}8 (purple) and \textsc{Herwig}7 (blue), while dashed lines show the fits including a sextic contribution according to \eqref{eq:sextic_fit} for \textsc{Pythia}8 (magenta) and \textsc{Herwig}7 (cyan).}
\label{fig:K_deltar_last5bins_pp_sextic}
\end{figure}

Following the discussion in Section~\ref{sec:analytical},  $K$ is expected to depend on powers of $(R_L/R)^{2n}$,  with the leading non-vanishing contribution appearing at order $n=1$ or higher. To test this behavior, we extract the values of $K(R_L)$ by performing linear fits to the data in Figs.~\ref{fig:Edge_deltaphi_last5bins_pp} and \ref{fig:Edge_deltaphi_last5bins_PbPb}, this time fixing $K_0 = 0$. These $K$-values along with their uncertainties are given in Appendix~\ref{appA} (see Tables~\ref{tab:fits_binbybin_pp_K0zero} and~\ref{tab:fits_binbybin_PbPb_K0zero} for the p-p and Pb-Pb results, respectively). We then plot $K$ as a function of $R_{L}/R$ and fit it with
\begin{equation}
K = a_2 \left(\frac{R_L}{R}\right)^2 + a_4 \left(\frac{R_L}{R}\right)^4  \,,
\label{eq:quartic_fit}
\end{equation}
where $a_2$ and $a_4$ are the fit parameters. Since edge effects manifest as a suppression in the EEC spectrum, the first non-zero coefficient in~\eqref{eq:quartic_fit} must be positive.

\begin{figure}[t]
\includegraphics[scale=0.47]{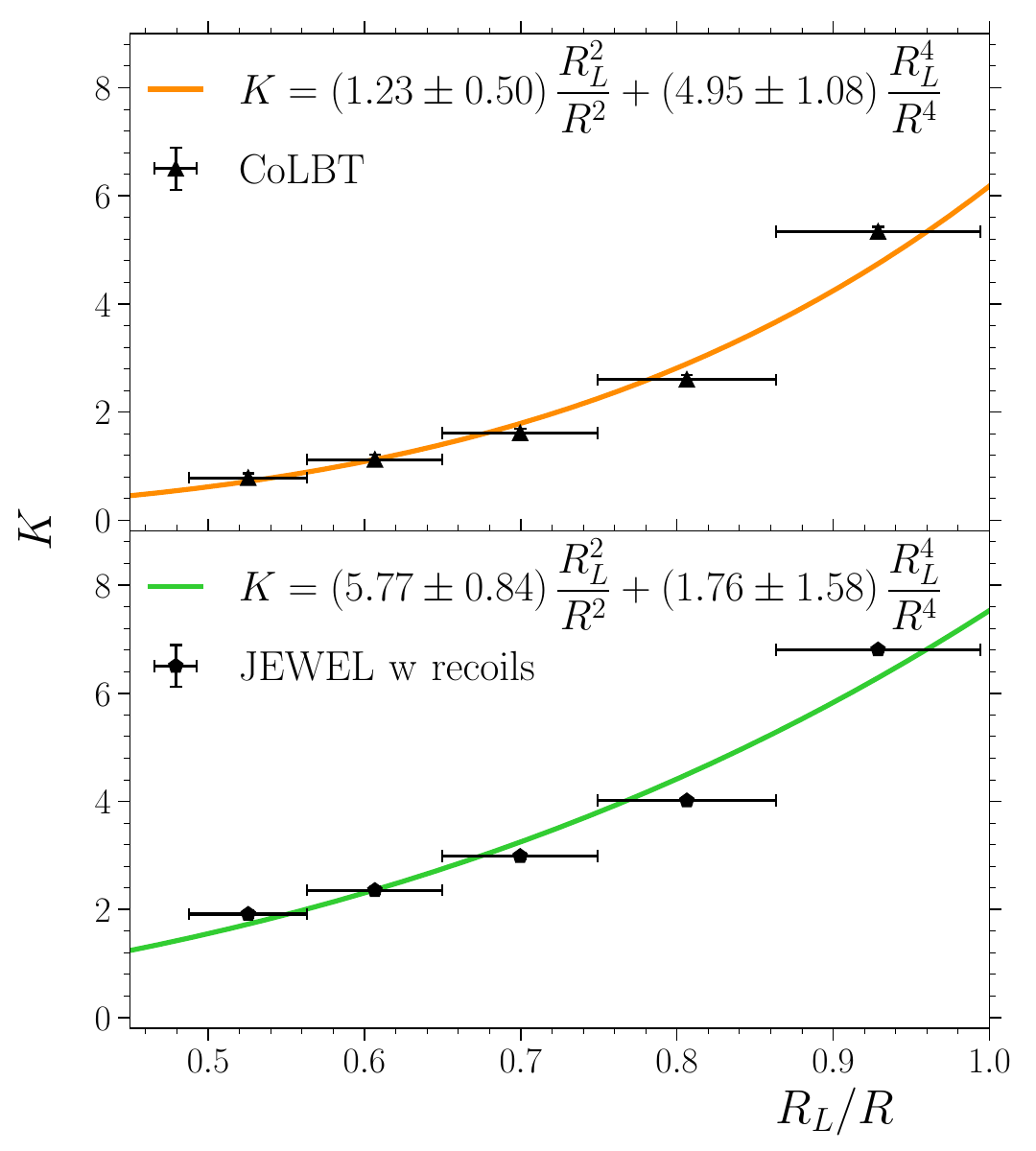}
\caption{Top: fitted values of $K$  for the highest five $R_L$ bins as a function of $R_L/R$ (with $R=0.4$)  for \textsc{CoLBT}  in $\sqrt{s_{\rm NN}} = 5.02$ TeV  0-10$\%$ Pb-Pb collisions. Bottom: same as top but for \textsc{JEWEL} with recoils. Solid lines show fits  following~\eqref{eq:quartic_fit} for  \textsc{CoLBT} (orange) and \textsc{JEWEL} with recoils (green).}
\label{fig:K_deltar_last5bins_PbPb}
\end{figure}

Fig.~\ref{fig:K_deltar_last5bins_pp_sextic} shows the \textsc{Pythia}8 (top) and \textsc{Herwig}7 (bottom) results for 
$K$ as a function of $R_L/R$ for $R=0.4$, together with their corresponding fits based on~\eqref{eq:quartic_fit} shown as solid lines. The fits provide a good overall description of both the \textsc{Pythia}8 and \textsc{Herwig}7 data. Notably, in both cases the fitted $a_2$ values are close to zero, indicating that in these MC generators the first significant contribution to the edge effects arises at order 
$(R_L /R)^4$. Building on this result, we also perform additional fits including a sextic term:
\begin{equation}
K = a_4 \left(\frac{R_L}{R}\right)^4 + a_6 \left(\frac{R_L}{R}\right)^6 \,,
\label{eq:sextic_fit}
\end{equation}
where $a_6$ captures the first higher-order correction beyond the quartic term. The resulting curves are shown as dashed lines in Fig.~\ref{fig:K_deltar_last5bins_pp_sextic}, for \textsc{Pythia}8 (magenta) and \textsc{Herwig}7 (cyan). Including the sextic term does not improve the description of the results, indicating   that the quartic term dominates the edge effects.

\begin{figure}[t]
\includegraphics[scale=0.47]{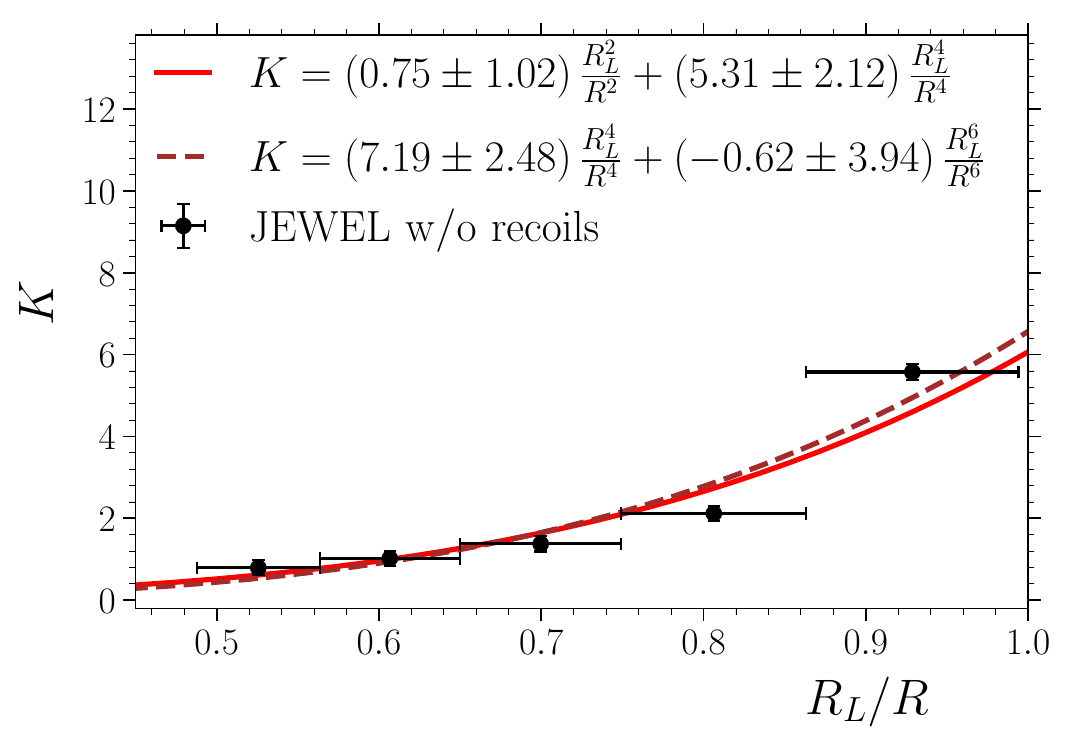}
\caption{Fitted values of $K$  for the highest five $R_L$ bins as a function of $R_L/R$ (with $R=0.4$)  for \textsc{JEWEL} without recoils in $\sqrt{s_{\rm NN}} = 5.02$ TeV 0-10$\%$ Pb-Pb collisions. The solid red curve shows the  fit following~\eqref{eq:quartic_fit}, while the dashed brown shows the fit including a sextic contribution according to \eqref{eq:sextic_fit}.}
\label{fig:K_deltar_last5bins_PbPb_JewelNo}
\end{figure}

Fig.~\ref{fig:K_deltar_last5bins_PbPb} shows the heavy-ion results. The $K$ values, extracted for \textsc{CoLBT} (top) and \textsc{JEWEL} with recoils (bottom), correspond to those listed in Table~\ref{tab:fits_binbybin_PbPb_K0zero} and are plotted as a function of $R_L/R$, together with fits based on~\eqref{eq:quartic_fit}. Unlike the p-p Monte Carlo results, where the quadratic term was negligible, the leading contribution to edge effects in these heavy-ion fits scales as $(R_L/R)^2$. This quadratic term is particularly enhanced in \textsc{JEWEL} with recoils, where the fitted $a_2$ is significantly larger than in \textsc{CoLBT} (see figure labels). Within uncertainties, the quartic coefficient $a_4$ for \textsc{JEWEL} (with recoils) is close to zero, indicating that, in this model, edge effects in the EEC for $R_L <0.4$ are largely accounted for by the leading $(R_L/R)^2$ term.

Up to now, all  \textsc{JEWEL} results include medium response via recoils. The same analysis can also be performed for \textsc{JEWEL} without recoils. Results for the  EEC($R=0.4$)/EEC($R=0.8$) ratio for several jet $p_T$ bins, together with the linear fits of the edge effects as a function of $\langle \phi \rangle$, are provided in the Appendix~\ref{appB}. Here, Fig.~\ref{fig:K_deltar_last5bins_PbPb_JewelNo} shows only the final fits for $K(R_L/R)$ in \textsc{JEWEL} without recoils. These $K$ values, obtained from the linear fits with $K_0$ set to zero, are listed in Table~\ref{tab:fits_binbybin_JewelNo_K0zero} in Appendix~\ref{appB}. The solid red curve,  following~\eqref{eq:quartic_fit}, shows that in  the absence of recoils, the leading contribution to edge effects in \textsc{JEWEL} enters at order $(R_L/R)^4$. This is in contrast with  \textsc{JEWEL} with recoils (bottom panel of Fig.~\ref{fig:K_deltar_last5bins_PbPb}), which exhibits  a clearly nonzero quadratic term, indicating that including recoils induces an $(R_L/R)^2$ enhancement of the edge effects. This enhanced scaling when in the presence of recoils is consistent with the enhancement of sub-leading power corrections due to nuclear modifications to the EEC expected from the light-ray OPE \cite{Andres:2024xvk}.
Finally, a sextic fit to the $K$ values in \textsc{JEWEL} without recoils, following~\eqref{eq:sextic_fit} (dashed line), shows that the sextic coefficient is negligible just as in vacuum. This indicates that for this \textsc{JEWEL} setup  edge effects in the EEC are well captured by the leading $(R_L/R)^4$ term for $R_L < 0.4$.

\begin{figure}[t]
\includegraphics[scale=0.46]{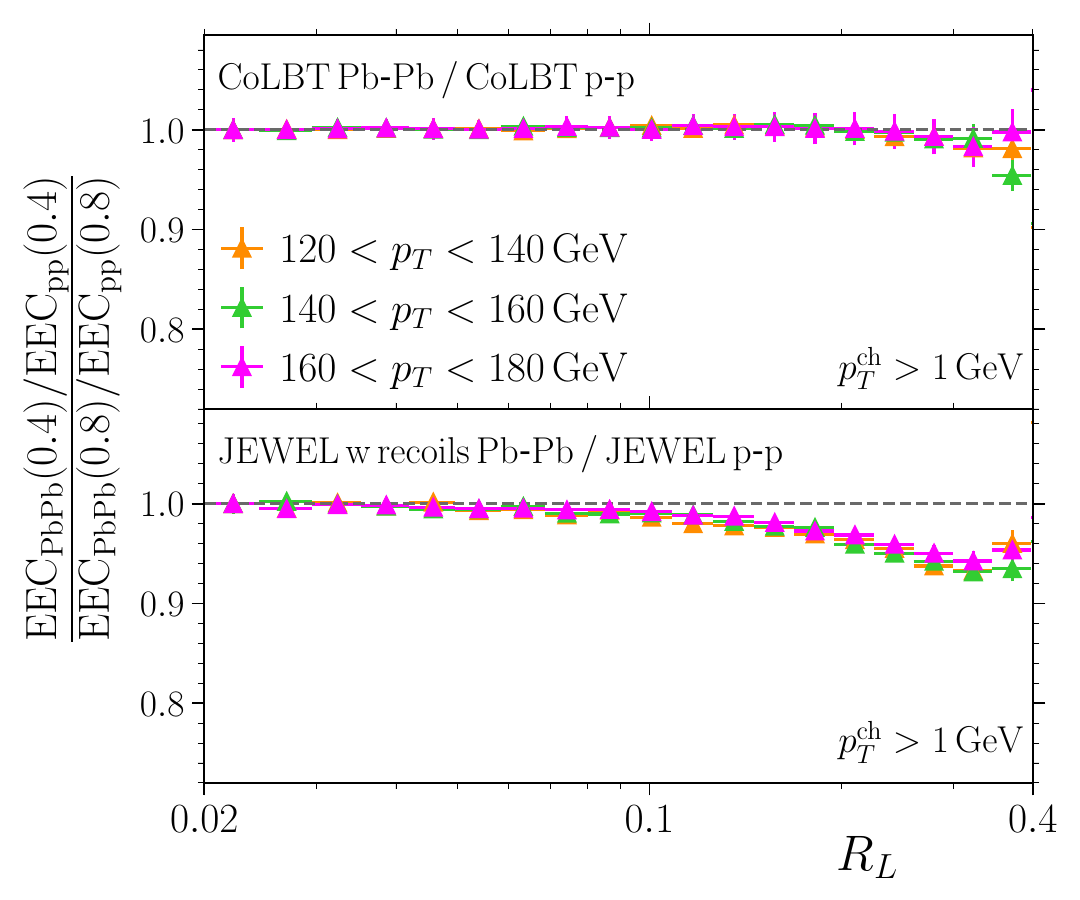}
\caption{Top: double ratio of the EEC, defined as the ratio of EEC($R=0.4$)/EEC($R=0.8$) in 0–10\%  $\sqrt{s_{\rm NN}} = 5.02$ TeV Pb-Pb collisions to the same ratio for its p-p reference, computed with \textsc{CoLBT}. Bottom: Same as the top panel, but using \textsc{JEWEL} with recoils.}
\label{fig:E2C_doubleratio}
\end{figure}

We now examine the double ratios of Pb-Pb/p-p for the EEC with $R=0.4$  relative to that with $R=0.8$ to assess the role of edge effects in the Pb-Pb/p-p ratio. The results for the \textsc{CoLBT} and \textsc{JEWEL} with recoils are shown in the top and bottom panels of Fig.~\ref{fig:E2C_doubleratio}, respectively, while the corresponding results for \textsc{JEWEL} without recoils  are provided in Appendix~\ref{appB}. We note that each double ratio is constructed for each heavy-ion MC using its own p-p  baseline. We find that edge effects are significantly reduced in the Pb-Pb/p-p ratio, although not entirely eliminated. This behavior is consistent with Eq.~\eqref{eq:doubleratio}, where edge effects in the Pb-Pb/p-p ratio  scale with $K^{\rm PbPb}-K^{\rm pp}$. Since, for any collision system, $K$ is strictly positive, the magnitude of the edge effects is always be smaller in the Pb-Pb/p-p ratio than in the individual spectra. However, because $K$ depends on the collision system (see Figs.~\ref{fig:K_deltar_last5bins_pp_sextic} and \ref{fig:K_deltar_last5bins_PbPb}), the cancellation is not perfect. Consequently, the residual edge effects in the Pb-Pb/p-p ratio are at most $\sim 8\%$, reduced from up to $25\%$ in the Pb-Pb EEC. The cancellation is particularly clean in CoLBT, consistent with the observation that in Pb-Pb in the \textsc{CoLBT} the quadratic term is not very large, while in its p-p baseline, edge effects scale as $R_L^{\, 4}$.

\begin{figure}[th]
\includegraphics[scale=0.47]{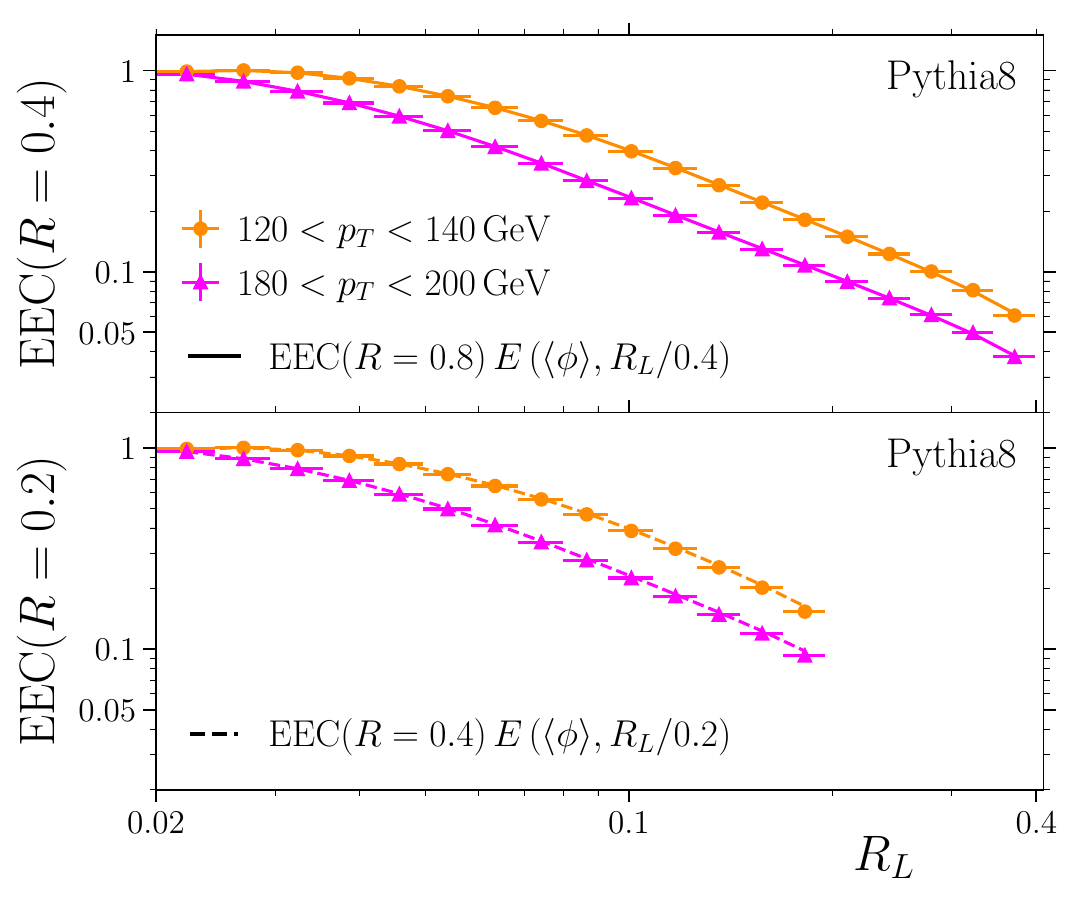}
\caption{Top: the markers show the EEC for jets with $120 < p_T < 140~{\rm GeV}$ (orange) and $180 < p_T < 200~{\rm GeV}$ (magenta) for jet radius $R=0.4$. The lines correspond to the EEC for $R=0.8$, multiplied by the factor defined in Eq.~\eqref{eq:edge_factor}, for each $p_T$ interval.
Bottom: same as top, but markers refer to $R=0.2$ jets, and dashed lines to the EEC for $R=0.4$ multiplied by Eq.~\eqref{eq:edge_factor}.
All results were obtained from the \textsc{Pythia}8 $\sqrt{s} = 5.02$ p-p samples.}
\label{fig:E2C_0p2_0p4_Pythia8}
\end{figure}

\begin{figure}[th]
\includegraphics[scale=0.47]{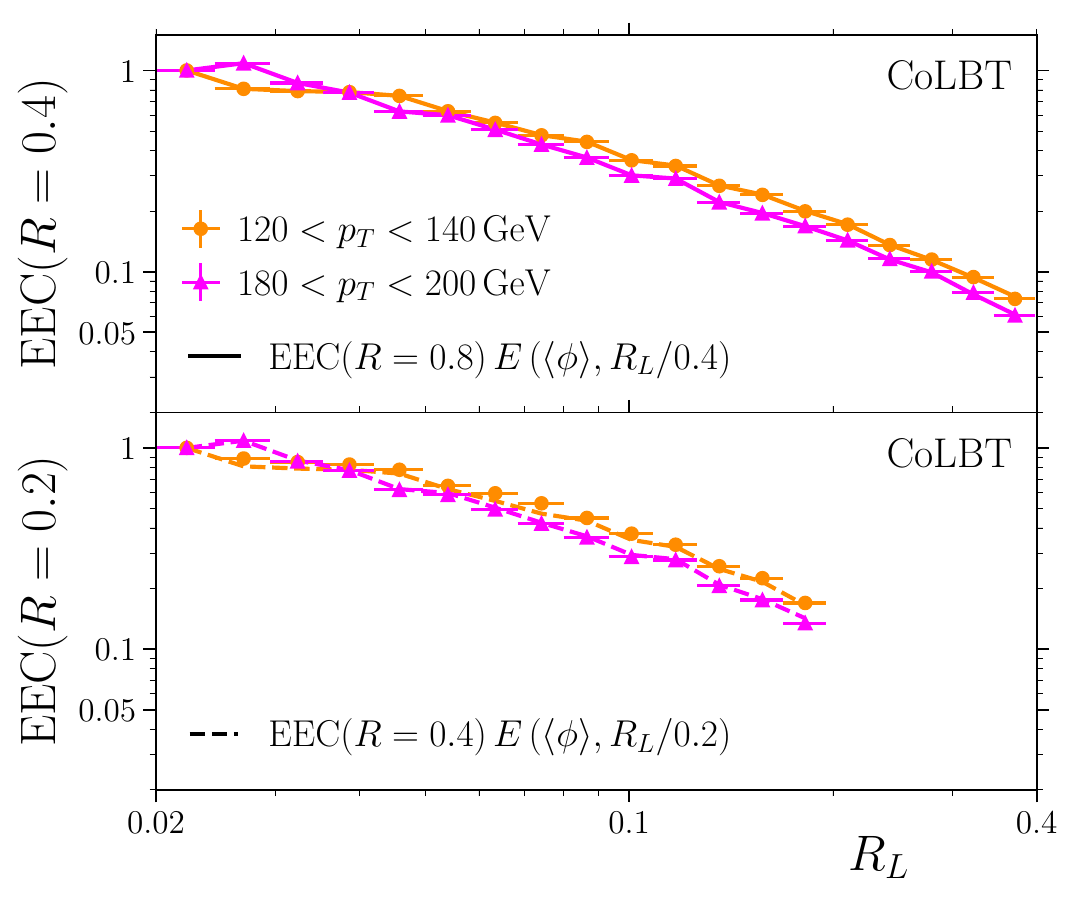}
\caption{Top: the markers show the EEC for jets with $120 < p_T < 140~{\rm GeV}$ (orange) and $180 < p_T < 200~{\rm GeV}$ (magenta) for jet radius $R=0.4$. The lines correspond to the EEC for $R=0.8$, multiplied by the factor defined in Eq.~\eqref{eq:edge_factor}, for each $p_T$ interval.
Bottom: same as top, but  markers refer to $R=0.2$ jets, and dashed to the EEC for $R=0.4$ multiplied by the factor in~\eqref{eq:edge_factor}.
All results were obtained using \textsc{CoLBT} Pb-Pb simulations at $\sqrt{s_{\rm NN}} = 5.02$ TeV in the  0-10$\%$ centrality class.}
\label{fig:E2C_0p2_0p4_CoLBT}
\end{figure}

Finally, we test how well the edge-effect model developed in this manuscript reproduces the modifications to the EEC spectrum induced by a finite jet radius. Figs.~\ref{fig:E2C_0p2_0p4_Pythia8} and \ref{fig:E2C_0p2_0p4_CoLBT} show representative comparisons for \textsc{Pythia}8 and \textsc{CoLBT}, respectively. In each figure, the markers correspond to the EEC obtained with smaller jet radii ($R=0.4$ in the upper panels and $R=0.2$ in the lower panels), shown for two representative $p_T$ bins: $120 < p_T < 140$,GeV (orange) and $180 < p_T < 200$,GeV (magenta). The lines show the corresponding EEC for the larger reference radius ($R=0.8$ in the upper panels and $R=0.4$ in the lower panels), multiplied by the edge-effect factor $E$ defined in Eq.~\eqref{eq:edge_factor}. In this set-up, the larger-radius spectrum therefore serves as a proxy for an edge-effect-free measurement within the angular range covered by the smaller radius, and the model prediction (solid line) is obtained by multiplying  this reference spectrum  by the extracted edge-effect factor $E$.
The same functional form of $K(R_L/R)$ is used to compute $E$ for all four curves in both figures, with $K$ given by the fit in Eq.~\eqref{eq:quartic_fit}. The coefficients $a_2$ and $a_4$  for all curves in Fig.~\ref{fig:E2C_0p2_0p4_Pythia8}  are those extracted for \textsc{Pythia}8 and shown in Fig.~\ref{fig:K_deltar_last5bins_pp_sextic}, , while the values used in Fig.~\ref{fig:E2C_0p2_0p4_CoLBT} correspond to \textsc{CoLBT} and are listed in Fig.~\ref{fig:K_deltar_last5bins_PbPb}.

As expected, the magnitude of edge effects varies with jet $p_T$, and the strong agreement between the data points and the model lines shows that this dependence is well captured through the single parameter $\langle \phi \rangle$. Importantly, the lower panels of both Figs.~\ref{fig:E2C_0p2_0p4_Pythia8} and \ref{fig:E2C_0p2_0p4_CoLBT} correspond to genuine predictions, as the $K$ functions used there were extracted from the $R \! = \! 0.4$/$R \! = \! 0.8$ comparison in \textsc{Pythia}8 and \textsc{CoLBT}, respectively. The observed agreement between the markers and the model curves indicates that our simple analytical model captures the suppression induced by edge effects across different jet radii and $p_T$ ranges.

%%%%%%%%%%%%%%%%%%%%%%%%%%%%%%%%%%%%%%%%%
\section{Conclusions and outlook}
\label{sec:conclusions}

In this work, we have developed an understanding of edge effects in soft-insensitive angular jet substructure observables, using the two-point energy correlator as a concrete case study. Central to our approach is the introduction of a phenomenological model that captures how these effects arise from the jet reconstruction procedure and propagate into the measured correlator spectra. This model predicts a simple scaling behavior: edge effects grow linearly with the average angular separation $\langle \phi \rangle$ between the winner-take-all and $E$-scheme jet axes. We validated this prediction in p-p simulations with \textsc{Pythia}8 and \textsc{Herwig}7, as well as in Pb-Pb simulations using \textsc{JEWEL} and \textsc{CoLBT}.

Our analysis reveals that in p-p collisions, the dominant contribution to edge effects scales as $(R_L/R)^4$, confirming that  these distortions are parametrically suppressed for sufficiently large jet radii. In contrast,  Pb-Pb simulations exhibit  substantially larger distortions, with \textsc{JEWEL} including recoils showing a significant $(R_L/R)^2$ component. This enhanced scaling is consistent with prior observations that the nuclear medium amplifies subleading power corrections in jet substructure observables \cite{Andres:2024xvk}. We further show that in Pb-Pb/p-p EEC ratios, edge effects are substantially reduced (typically below the $\lesssim 10\%$ level) but do not fully cancel. This underscores the importance of including or constraining edge effects when interpreting heavy-ion jet substructure measurements.

A key motivation for this study is the uniquely high level of analytical control afforded by energy correlators, which can be computed from first principles using the light-ray OPE \cite{Hofman:2008ar}, even in the presence of a nuclear medium \cite{Andres:2024xvk}. While this approach has successfully described the angular and $p_T$-dependence of the modifications observed in the perturbative regime of the EEC measurements in  Pb-Pb  relative to p-p \cite{Andres:2024xvk}, it does not account for edge effects, complicating the extraction of underlying in-medium matrix elements from experimental data. The phenomenological framework developed here provides a systematic method to quantify the theoretical uncertainty associated with neglecting edge effects in the extraction of these matrix elements, thereby mitigating the distortions introduced by edge effects in theory-experiment comparisons. 

One of the broader insights of this study is that the magnitude of edge effects in angular substructure observables is largely dictated by the $\langle \phi \rangle$ distribution of jet axis decorrelation. This places a new emphasis on accurately modeling this distribution in event generators. While generators naturally include edge effects, their predictions in the heavy-ion environment are difficult to benchmark due to the lack of analytic control. By demonstrating that the leading distortions are governed by $\langle \phi \rangle$, we provide a direct path for validation: the $\langle \phi \rangle$ distribution is analytically calculable in vacuum \cite{Cal:2019gxa} and is also experimentally measurable in A-A and p-p collisions \cite{ALICE:2022rdg,ALICE:2023dwg,CMS:2025dnx,CMS:2024msa}, enabling cross-checks between theory, generators, and data that were previously inaccessible. This establishes a new avenue for assessing the modeling fidelity of heavy-ion event generators and strengthens the robustness of jet-based probes of quark–gluon plasma dynamics.

A final remark about our findings. Since QGP-induced modifications to the EEC scale as $R_L^2$ relative to vacuum, according to the light-ray OPE, measurements of correlator-based observables in heavy-ion jets with larger radii could provide complementary insight to QGP-induced phenomena, wherein edge effects will be manifest differently. However, such measurements remain extremely challenging in Pb–Pb collisions due to the large background. Highly energetic light-ion collisions \cite{Brewer:2021kiv}, such as the recent oxygen-oxygen and neon-neon runs at the LHC, offer a promising avenue, where the much smaller background renders these large-$R$ measurements in principle feasible, thereby opening a wider kinematic regime in which the analytical control of the OPE can be most effectively leveraged.

%%%%%%%%%%%%%%%%%%%%%%%%%%%%%%%%%%%%%%%%%
\section*{Acknowledgments}
We thank Philipp Aretz, Luna Chen, Arjun Kudinoor, Kyle Lee, Cyrille Marquet, Ian Moult, Jean-François Paquet, Jean du Plessis, Krishna Rajagopal and Rachel Steinhorst for useful discussions. AC and JH thank the gracious hospitality of the VandyGRAF group at Vanderbilt University. This work is funded by the European Union (ERC, QGPthroughEECs,  grant agreement No. 101164102). Views and opinions expressed are however those of the authors only and do not necessarily reflect those of the European Union or the European Research Council. Neither the European Union nor the granting authority can be held responsible for them. The authors also appreciate the support of the Vanderbilt ACCRE computing facility. RKE acknowledges funding from the U.S. Department of Energy, Office of Science, Office of Nuclear Physics under grant DE-SC0024660. JV would like to acknowledge funding by the U.S. Department of Energy, under grant number DE-FG05-92ER40712. ZY and BK acknowledge internal funding from Vanderbilt University. The work of CA is also partially supported by the U.S. Department of Energy, Office of Science, Office of Nuclear Physics under grant Contract Number DE-SC0011090. JH is supported by the Leverhulme Trust as an Early Career Fellow. 

%%
%%
%%%%%%%%%%% Appendix A %%%%%%%%%
\appendix
\section{Fit Parameters}
\label{appA}
For completeness, we summarize in this appendix the values of the parameters obtained from linear fits of the  edge effects differential in $R_L$, $\Delta_{{\rm edge \, effects}}$,  as a function of the average deflection between the WTA and $E$-scheme axes $\langle \phi \rangle$. The fits follow the functional form introduced in Eq.~\eqref{eq:linear_fit}. Table~\ref{tab:fits_binbybin_pp} lists the results for \textsc{Pythia}8 and \textsc{Herwig}7, corresponding to the fits shown in Fig.~\ref{fig:Edge_deltaphi_last5bins_pp}, while Table~\ref{tab:fits_binbybin_PbPb} presents the results for \textsc{CoLBT} and \textsc{JEWEL} with recoils, corresponding to the fits shown in Fig.~\ref{fig:Edge_deltaphi_last5bins_PbPb}. Tables \ref{tab:fits_binbybin_pp_K0zero} and \ref{tab:fits_binbybin_PbPb_K0zero} show the same fits but with the intercept $K_0$ fixed to zero. The $K$ values reported in Tables~\ref{tab:fits_binbybin_pp_K0zero} and~\ref{tab:fits_binbybin_PbPb_K0zero} are those plotted in Fig.~\ref{fig:K_deltar_last5bins_pp_sextic} and Fig.~\ref{fig:K_deltar_last5bins_PbPb}, respectively.
%%
%%%% Bin by bin: pp
\begin{table}[h]
\centering
\renewcommand{\arraystretch}{1.4} % adjust row height
\begin{tabular}{|c|c|c|c|}
    \hline
    & & $\boldsymbol{K_0}$ & $\boldsymbol{K}$ \\
    \hline
    \multirow{5}{*}{{\bf \textsc{Pythia}8}} 
        & $\boldsymbol{ R_L = 0.21}$ & 0.0067 $\pm$ 0.0021  & 0.200 $\pm$ 0.053 \\
        & $\boldsymbol{R_L = 0.24}$ & 0.0011 $\pm$ 0.0020  & 0.499 $\pm$ 0.053\\
        & $\boldsymbol{ R_L = 0.28}$ & -0.0088 $\pm$ 0.0020   & 1.027 $\pm$ 0.052 \\
        & $\boldsymbol{ R_L = 0.32}$ & -0.0297 $\pm$ 0.0020  & 2.171 $\pm$ 0.052 \\
        & $\boldsymbol{R_L = 0.37}$ & -0.0529 $\pm$ 0.0020 & 4.852 $\pm$ 0.050\\
    \hline
    \multirow{5}{*}{{\bf \textsc{Herwig}7}} & 
        $\boldsymbol{ R_L = 0.21}$ & 0.0019 $\pm$ 0.0065  &  0.378 $\pm$ 0.015 \\
        & $\boldsymbol{ R_L = 0.24}$ & -0.0061 $\pm$ 0.0065  & 0.706 $\pm$ 0.015\\
        & $\boldsymbol{ R_L = 0.28}$ & -0.0183 $\pm$ 0.0065 &  1.235 $\pm$ 0.014 \\
        & $\boldsymbol{ R_L = 0.32}$ & -0.0388 $\pm$ 0.0062   & 2.244 $\pm$ 0.014 \\
        & $\boldsymbol{ R_L = 0.37}$  & -0.067 $\pm$ 0.0062 & 4.594   $\pm$ 0.014 \\
    \hline
\end{tabular}
\caption{Parameters corresponding to the linear fit in \eqref{eq:linear_fit} for the \textsc{Pythia}8 and \textsc{Herwig}7 curves  shown in Fig.~\ref{fig:Edge_deltaphi_last5bins_pp}.}
\label{tab:fits_binbybin_pp}
\end{table}

%%%% Bin by bin: PbPb
\begin{table}[h]
\centering
\renewcommand{\arraystretch}{1.4} % adjust row height
\begin{tabular}{|c|c|c|c|}
    \hline
    & & $\boldsymbol{K_0}$ & $\boldsymbol{K}$ \\
    \hline
    \multirow{5}{*}{{\bf CoLBT}} 
        & $\boldsymbol{ R_L = 0.21}$ &  -0.0016 $\pm$ 0.0015 &   0.82 $\pm$  0.66\\
        & $\boldsymbol{ R_L = 0.24}$ & -0.017 $\pm$ 0.012  &  1.54 $\pm$ 0.89 \\
        & $\boldsymbol{ R_L = 0.28}$ & -0.016 $\pm$  0.012 &  2.00 $\pm$ 0.90  \\
        & $\boldsymbol{ R_L = 0.32}$ & -0.036 $\pm$ 0.013  & 3.50 $\pm$ 0.97 \\
        & $\boldsymbol{ R_L = 0.37}$ & 0.011 $\pm$ 0.013 &  5.07 $\pm$ 0.97 \\
    \hline
    \multirow{5}{*}{ \makecell{\textbf{\textsc{JEWEL} }\\\textbf{with recoils}}} & 
        $\boldsymbol{ R_L = 0.21}$ & -0.023 $\pm$ 0.013 &  2.63 $\pm$ 0.41 \\
        & $\boldsymbol{ R_L = 0.24}$ & -0.020 $\pm$ 0.012  & 3.01 $\pm$ 0.38 \\
        & $\boldsymbol{ R_L = 0.28}$ & -0.022 $\pm$  0.011 & 3.67 $\pm$ 0.35 \\
        & $\boldsymbol{ R_L = 0.32}$ & -0.023 $\pm$  0.010  &  4.72 $\pm$ 0.32 \\
        & $\boldsymbol{ R_L = 0.37}$  & 0.0042 $\pm$ 0.0043 &  6.67 $\pm$  0.30 \\
    \hline
\end{tabular}
\caption{Parameters corresponding to the linear fit in \eqref{eq:linear_fit} for the \textsc{CoLBT} and \textsc{JEWEL} with recoils curves shown in Fig.~\ref{fig:Edge_deltaphi_last5bins_PbPb}.}
\label{tab:fits_binbybin_PbPb}
\end{table}
%%
%%
%%%% Bin by bin: pp
\begin{table}[th]
\centering
\renewcommand{\arraystretch}{1.4} % adjust row height
\begin{tabular}{|c|c|c|c|}
    \hline
    & & $\boldsymbol{K}$ \\
    \hline
    \multirow{5}{*}{{\bf \textsc{Pythia}8}} 
        & $\boldsymbol{ R_L = 0.21}$ &  0.3748  $\pm$ 0.0078 \\
        & $\boldsymbol{R_L = 0.24}$  &0.5278  $\pm$ 0.0077\\
        & $\boldsymbol{ R_L = 0.28}$ & 0.8024$\pm$ 0.0077 \\
        & $\boldsymbol{ R_L = 0.32}$ & 1.4169 $\pm$ 0.0076  \\
        & $\boldsymbol{R_L = 0.37}$ & 3.5148 $\pm$ 0.0072\\
    \hline
    \multirow{5}{*}{{\bf \textsc{Herwig}7}} & 
        $\boldsymbol{ R_L = 0.21}$ & 0.423  $\pm$ 0.019 \\
        & $\boldsymbol{ R_L = 0.24}$  & 0.570  $\pm$ 0.018\\
        & $\boldsymbol{ R_L = 0.28}$ &  0.828 $\pm$ 0.018 \\
        & $\boldsymbol{ R_L = 0.32}$ & 1.381 $\pm$ 0.017  \\
        & $\boldsymbol{ R_L = 0.37}$  &  3.116  $\pm$ 0.017  \\
    \hline
\end{tabular}
\caption{Values of  $K$ obtained by  fitting the \textsc{Pythia}8 and \textsc{Herwig}7 $\Delta_{{\rm edge \, effects}}$ data shown in Fig.~\ref{fig:Edge_deltaphi_last5bins_pp} to Eq.~\eqref{eq:linear_fit} with $K_0=0$. 
These  $K$ values correspond to the points plotted in Fig.~\ref{fig:K_deltar_last5bins_pp_sextic}.}
\label{tab:fits_binbybin_pp_K0zero}
\end{table}

%%%% Bin by bin: PbPb. K0=0
\begin{table}[h]
\centering
\renewcommand{\arraystretch}{1.4} % adjust row height
\begin{tabular}{|c|c|c|c|}
    \hline
    &  & $\boldsymbol{K}$ \\
    \hline
    \multirow{5}{*}{{\bf CoLBT}} 
        & $\boldsymbol{ R_L = 0.21}$  & 0.784 $\pm$  0.083\\
        & $\boldsymbol{ R_L = 0.24}$  & 1.120 $\pm$ 0.085 \\
        & $\boldsymbol{ R_L = 0.28}$  & 1.612 $\pm$ 0.085  \\
        & $\boldsymbol{ R_L = 0.32}$  & 2.600 $\pm$   0.090\\
        & $\boldsymbol{ R_L = 0.37}$  & 5.338 $\pm$  0.088\\
    \hline
    \multirow{5}{*}{ \makecell{\textbf{\textsc{JEWEL} }\\\textbf{with recoils}}} & 
        $\boldsymbol{ R_L = 0.21}$  &  1.9161  $\pm$ 0.077  \\
        & $\boldsymbol{ R_L = 0.24}$ & 2.3565 $\pm$  0.071\\
        & $\boldsymbol{ R_L = 0.28}$  & 2.9868  $\pm$  0.065\\
        & $\boldsymbol{ R_L = 0.32}$ &  4.0139 $\pm$ 0.059\\
        & $\boldsymbol{ R_L = 0.37}$  & 6.8062 $\pm$  0.056  \\
    \hline
\end{tabular}
\caption{Values of  $K$ obtained by fitting  the \textsc{CoLBT}  and \textsc{JEWEL} with recoils $\Delta_{{\rm edge \, effects}}$ data shown in Fig.~\ref{fig:Edge_deltaphi_last5bins_PbPb} to Eq.~\eqref{eq:linear_fit} with $K_0=0$. 
These  $K$ values correspond to the points plotted in Fig.~\ref{fig:K_deltar_last5bins_PbPb}.}
\label{tab:fits_binbybin_PbPb_K0zero}
\end{table}
%%
%%

%%%%%%%%%%%%%%%%%%%%%%%%%%%%%%%%%
%%%%%%%%%%% Appendix B %%%%%%%%%
\section{\textsc{JEWEL} without recoils}
\label{appB}

\begin{figure}
\includegraphics[scale=0.47]{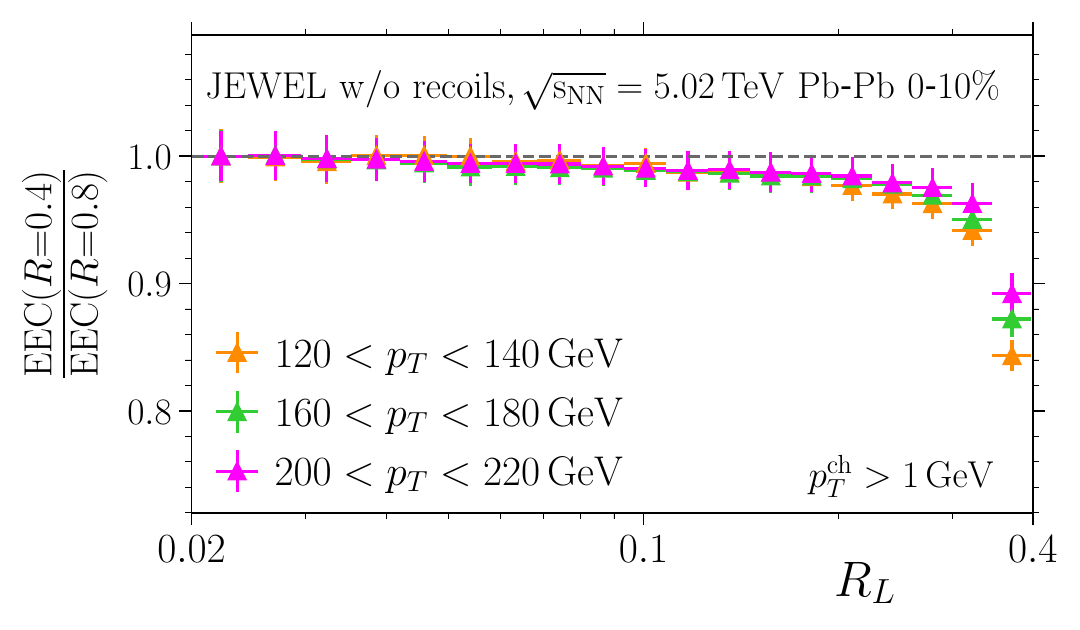}
\caption{Ratio of the EEC computed within anti-$k_T$ jets with radius $R = 0.4$ to that obtained with $R = 0.8$,  using the same sample of events generated with \textsc{JEWEL} without recoils in $\sqrt{s_{\rm NN}} = 5.02$ TeV 0-10$\%$ Pb-Pb collisions, shown for three different jet $p_T$ bins.}
\label{fig:EEC_ratio_jewelNo}
\end{figure}

We present in this appendix complementary results obtained using \textsc{JEWEL} without recoils.   As in the main text, we use inclusive
jet samples generated with \textsc{JEWEL} v$2.4$ in $\sqrt{s_{\rm NN}} = 5.02$ TeV 0-10$\%$ Pb-Pb collisions. Fig.~\ref{fig:EEC_ratio_jewelNo} shows the ratio of the EEC for jets with radius $R=0.4$ to those with $R=0.8$ for three $p_T$ bins. Deviations from unity at large $R_L$ indicate that edge effects become more pronounced for lower jet $p_T$, as expected.

\begin{figure}
\includegraphics[scale=0.47]{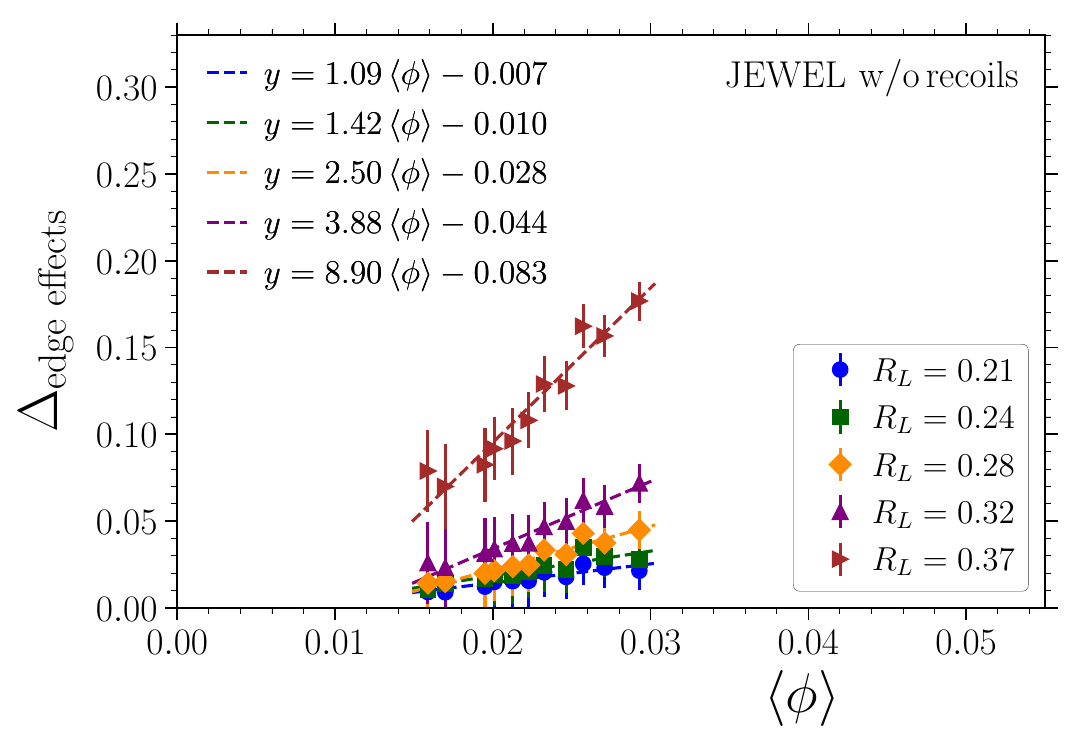}
\caption{Edge effects defined as in \eqref{eq:edge_effects} for the highest five $R_L$ bins, shown as a function of the average jet-axis difference  $\langle \phi \rangle$  for \textsc{JEWEL} without recoils  in $\sqrt{s_{\rm NN}} = 5.02$ TeV  0-10$\%$ Pb-Pb collisions. Linear fits for each $R_L$ bin are also included. The corresponding fit parameters, including their errors, are listed in Table~\ref{tab:fits_binbybin_JewelNo} of this appendix}
\label{fig:Edge_deltaphi_last5bins_JewelNo}
\end{figure}

Fig.~\ref{fig:Edge_deltaphi_last5bins_JewelNo} shows the edge effects, defined as in Eq.~\eqref{eq:edge_effects} of the main text, plotted as a function of the average angular separation between the WTA and $E$-scheme axes $\langle \phi \rangle$ for the highest five $R_L$ bins.  As for the \textsc{JEWEL} with recoils case in Fig.~\ref{fig:Edge_deltaphi_last5bins_PbPb}, the analysis without recoils  includes eleven  $p_T$ bins: $100 < p_T < 120$, $140 < p_T < 160$, …, $300 < p_T < 320$ GeV. The dashed lines correspond to linear fits of the MC results following~\eqref{eq:linear_fit}, showing that, as in the case with recoils, the edge effects in \textsc{JEWEL} without recoils grow approximately linearly with $\langle \phi \rangle$, with intercepts close to or compatible with zero. The resulting fit parameters, together with their uncertainties, are provided in Table~\ref{tab:fits_binbybin_JewelNo}. We also repeat the linear fits, but with the intercept $K_0$ fixed to zero. The corresponding $K$ values,  shown in Table~\ref{tab:fits_binbybin_JewelNo_K0zero}, are those plotted in Fig.~\ref{fig:K_deltar_last5bins_PbPb_JewelNo}.

%%%% \textsc{JEWEL} no recoils
\begin{table}
\centering
\renewcommand{\arraystretch}{1.4} % adjust row height
\begin{tabular}{|c|c|c|c|}
    \hline
    & & $\boldsymbol{K_0}$ & $\boldsymbol{K}$ \\
    \hline
    \multirow{5}{*}{ \makecell{\textbf{\textsc{JEWEL} }\\\textbf{without recoils}}} & 
        $\boldsymbol{ R_L = 0.21}$ & -0.0074 $\pm$ 0.0079  & 1.09 $\pm$  1.19 \\
        & $\boldsymbol{ R_L = 0.24}$ &   -0.010 $\pm$   0.012& 1.42 $\pm$ 1.20 \\
        & $\boldsymbol{ R_L = 0.28}$ &  -0.028 $\pm$ 0.028  &2.49 $\pm$ 1.24  \\
        & $\boldsymbol{ R_L = 0.32}$ & -0.044  $\pm$ 0.030     &3.88 $\pm$  1.23 \\
        & $\boldsymbol{ R_L = 0.37}$  &  -0.083  $\pm$ 0.031 &   8.90 $\pm$ 1.26   \\
    \hline
\end{tabular}
\caption{Parameters corresponding to the linear fit in Eq.~\eqref{eq:linear_fit}  for the  \textsc{JEWEL} without recoils curves shown in Fig.~\ref{fig:Edge_deltaphi_last5bins_JewelNo}.}
\label{tab:fits_binbybin_JewelNo}
\end{table}
%%
%%

%%%% \textsc{JEWEL} no recoils.k0=0
\begin{table}
\centering
\renewcommand{\arraystretch}{1.4} % adjust row height
\begin{tabular}{|c|c|c|}
    \hline
     &  & $\boldsymbol{K}$ \\
    \hline
    \multirow{5}{*}{\makecell{\textbf{JEWEL}\\\textbf{without recoils}}} 
        & $\boldsymbol{R_L=0.21}$ & $0.79  \pm 0.18$ \\
        & $\boldsymbol{R_L=0.24}$ & $1.02  \pm 0.18$ \\
        & $\boldsymbol{R_L=0.28}$  & $1.37 \pm 0.18$ \\
        & $\boldsymbol{R_L=0.32}$  & $2.12 \pm 0.19$ \\
        & $\boldsymbol{R_L=0.37}$  &5.57 $ \pm 0.19$ \\
    \hline
\end{tabular}
\caption{Values of $K$ obtained by fitting  the \textsc{JEWEL}  without recoils $\Delta_{{\rm edge \, effects}}$ data shown in Fig.~\ref{fig:Edge_deltaphi_last5bins_JewelNo} to Eq.~\eqref{eq:linear_fit} with $K_0 = 0$. These $K$ values correspond to those plotted in Fig.~\ref{fig:K_deltar_last5bins_PbPb_JewelNo}.}
\label{tab:fits_binbybin_JewelNo_K0zero}
\end{table}

\begin{figure}[th]
\includegraphics[scale=0.47]{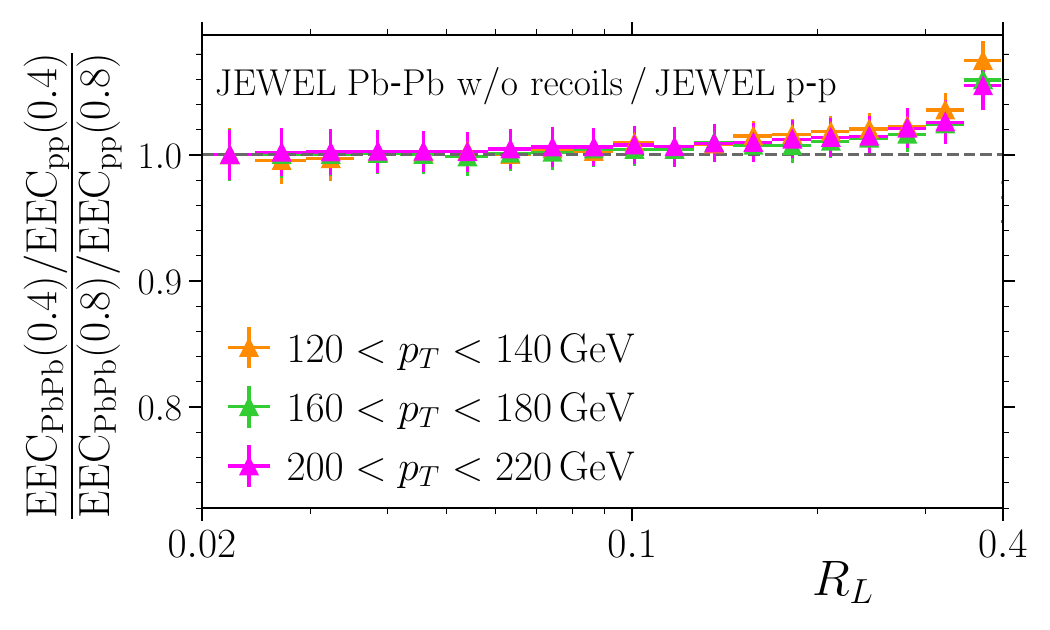}
\caption{Double ratio of the EEC, defined as the ratio of EEC($R=0.4$)/EEC($R=0.8$) in $\sqrt{s_{\rm NN}} = 5.02$ TeV 0-10$\%$ Pb-Pb collisions to the same ratio for its p-p reference, computed with  \textsc{JEWEL} without recoils.}
\label{fig:E2C_doubleratio_Jewelnorecoils}
\end{figure}

Finally, Fig.~\ref{fig:E2C_doubleratio_Jewelnorecoils} shows the double ratio of Pb-Pb/p-p for the $R=0.4$ EEC relative to the $R=0.8$ EEC  using \textsc{JEWEL} without recoils for the same three $p_T$-bins shown in Fig.~\ref{fig:EEC_ratio_jewelNo}. We note that the p-p EEC spectra entering this double ratio are taken from the vacuum \textsc{Pythia}6 baseline in JEWEL. As expected, edge effects are reduced in the double ratio compared to the corresponding Pb-Pb EEC ratio shown in Fig.~\ref{fig:EEC_ratio_jewelNo}, illustrating that the double ratio mitigates, although does not completely eliminate, the impact of finite-radius edge effects. We further note that this double ratio rises above 1 at large $R_L$, indicating that the suppression in the EEC spectrum due to edge effects is smaller in Pb-Pb than in p-p within \textsc{JEWEL} without recoils.

\bibliography{refs.bib}{}
\bibliographystyle{apsrev4-1}
\newpage
\onecolumngrid
\newpage

\end{document}